\documentclass[prx,aps,twocolumn,reprint,floatfix,showpacs,10pt]{revtex4-1}
\usepackage{amsmath}
\usepackage{graphicx}
\usepackage{epsfig}
\usepackage{graphics}
\usepackage{bm}
\usepackage{amssymb}
\usepackage{psfrag}
\usepackage{mathtools}
\usepackage[english]{babel}
\usepackage[T1]{fontenc}
\usepackage{lmodern}
\usepackage{color}
\usepackage{braket}
\usepackage{ulem}
\usepackage{booktabs}
\usepackage{siunitx}
\usepackage{xcolor}

\usepackage{umoline}

\usepackage[colorlinks,bookmarks=false,citecolor=blue,linkcolor=red,urlcolor=magenta]{hyperref}
\usepackage{amssymb,amsmath} 
\usepackage[T1]{fontenc}
\usepackage{epsfig}

\def\urlprefix{}
\def\url#1{}
\def\be{\begin{equation}}
\def\ee{\end{equation}}
\def\bea{\begin{eqnarray}}
\def\eea{\end{eqnarray}}
\def\bi{\begin{itemize}}
\def\ei{\end{itemize}}
\def\bin{\begin{enumerate}}
\def\ein{\end{enumerate}}

\def\bg{\begin{equation}\begin{gathered}}
\def\eg{\end{gathered}\end{equation}}

\def\bgg{\begin{equation*}\begin{gathered}}
\def\egg{\end{gathered}\end{equation*}}

\def\bgg{\begin{equation*}\begin{gathered}}
\def\egg{\end{gathered}\end{equation*}}

\begin{document}
\title{Level statistics across the many--body localization transition}

\author{Piotr Sierant}
\affiliation{
Instytut Fizyki imienia Mariana Smoluchowskiego, 
Uniwersytet Jagiello\'nski, ulica Profesora Stanis\l{}awa \L{}ojasiewicza 11, PL-30-348 Krak\'ow, Poland}
\author{Jakub Zakrzewski} 
\affiliation{
Instytut Fizyki imienia Mariana Smoluchowskiego, 
Uniwersytet Jagiello\'nski, ulica Profesora Stanis\l{}awa \L{}ojasiewicza 11, PL-30-348 Krak\'ow, Poland}
\affiliation{Mark Kac Complex Systems Research Center, Uniwersytet Jagiello\'nski, ulica Profesora Stanis\l{}awa \L{}ojasiewicza 11, PL-30-348 Krak\'ow, Poland}
\email{jakub.zakrzewski@uj.edu.pl}

\date{\today}

\begin{abstract}

Level statistics of systems that undergo many--body localization transition are studied. 
 An analysis of the gap ratio statistics
from the perspective of inter- and intra-sample randomness allows us to
pin point differences between transitions in random and quasi-random disorder, 
showing the effects due to  Griffiths rare events for the former case. 
It is argued that the transition in the case of random disorder 
exhibits universal features that are identified by constructing an appropriate 
model of intermediate spectral statistics which is a generalization
of the family of short-range plasma models.  
The considered weighted short-range plasma model yields a very good agreement both for 
level spacing distribution including its exponential tail and the number variance up to tens of level spacings
outperforming previously proposed models. 
In particular, our model grasps the critical level statistics which arise at disorder strength for 
which the inter-sample fluctuations are the strongest. Going beyond the paradigmatic examples of 
many-body localization in spin systems, we show
that the considered model also grasps the level statistics of disordered Bose- and Fermi-Hubbard models. 
The remaining deviations for long-range spectral correlations are discussed and attributed 
mainly to the intricacies of level unfolding.
\end{abstract}
\maketitle

\section{Introduction}
 It is 90 years already since Wishart in a seminal paper \cite{Wishart28} introduced the concept of
 random matrices into science. His original aim was to generalize the chi-squared distribution to 
 multiple dimensions, random symmetric non negative matrices played then the role of random variables. 
 The corresponding Wishart distribution found many applications from modern random matrix theory 
 \cite{Janik03} to various applications in physics \cite{Beenakker93,Verbaarschot93,Fyodorov97,Fyodorov99}, 
 wireless communications 
 \cite{Zanella09}  financial data for
large portfolios \cite{Bouchard01} etc. 

The next big step came with the introduction of Gaussian ensembles and the realization of Wigner
and others \cite{Porter} that  spectra for usually unknown complex nuclear Hamiltonians may be 
understood statistically  using properties of these ensembles obeying appropriate symmetries. 
It became a textbook knowledge that there exist exactly three universality classes \cite{Mehtabook,Haake} 
the Gaussian Orthogonal Ensemble (GOE) corresponds to systems invariant with respect to (generalized) time-reversal, 
the Gaussian Unitary Ensemble corresponds to systems with broken time-reversal invariance and the symplectic 
ensemble corresponds to half-integer spin systems with preserved time-reversal invariance and no other symmetries present. 
Thus, since the sixties it was the common knowledge that spectra of many-body interacting systems are statistically 
well described by random matrix theory (RMT). Further justifications of successes of RMT come from the theory of 
Dyson yielding the gaussian ensembles from an appropriate statistical mechanics description 
\cite{Dyson62a,Dyson62b,Dyson62c,Dyson72}.  
 
 An interesting development appeared in the eighties -- the conjecture that statistical properties of spectra 
 of systems chaotic in the classical limit are faithful to random matrix predictions \cite{Bohigas84}. This came 
 as a surprise - even simple single particle Hamiltonians 
 containing no randomness and represented by large, very sparse (due to strong selection rules in appropriately 
 chosen basis) matrices were statistically faithful to  RMT predictions as revealed e.g. in
 the study of hydrogen atom spectrum in the presence of strong magnetic field inducing the so called quadratic 
 Zeeman coupling \cite{Friedrich89}. More precisely, after unfolding the levels (obtaining the mean density of states equal to 
 unity) the remaining fluctuations were faithfully represented 
 by predictions of RMT \cite{Eckhardt88} as shown by nearest neighbor spacing distribution, $P(s)$, the so called
 number variance  (i.e. the variance of the number of levels in an interval of length $L$), correlation functions
 etc. The same measures indicated, however, that the transition from the chaotic to integrable situation (described 
 by Poisson ensemble of uncorrelated levels for systems of large dimensions \cite{Eckhardt88}) seems system specific
 and determined by the structure of the underlying classical mechanics in the mixed phase space  \cite{Bohigas93}.
 
 Similar transition from extended to localized states as revealed e.g. by a change of level
 statistics from GOE-like to Poisson-like appears in the Anderson localization transition.
 The corresponding level statistics has been addressed in the seminal paper \cite{Shklovskii93}
 followed by other important developments \cite{Kravtsov94,Aronov95,Kravtsov95} to mention early contributions --  
 for a review see \cite{Evers08}. In those cases a single particle problem in disordered medium was addressed.
 
 Recent years provided another important example of such a transition between ergodic (describable by standard
 gaussian RMT) and integrable limits -- the many-body localization (MBL). 
 While for weak disorder many-body interacting systems behave as expected for a long time being
 ergodic and following gaussian RMT predictions, for a sufficiently strong disorder a
 gradual (for finite systems sizes) crossover to localized situation occurs \cite{Basko06}. This phenomenon
 attracted an enormous interest in the last 10 years as it provides a robust example of non ergodic behavior in a
 complex many-body system. Instead of effectively thermalizing (as suggested by the eigenvector thermalization
 hypothesis (ETH) \cite{Srednicki94}) such a strongly disordered systems often remember their initial state as
 manifested in a series of spectacular experiments \cite{Schreiber15,Smith16,Choi16}. Already early theoretical studies 
 \cite{Oganesyan07} showed that a transition to MBL situation is accompanied by a change of level 
 statistics from that corresponding to 
 GOE to Poisson-like for MBL. 
 
 Importantly, it has been suggested that MBL phase is indeed integrable \cite{Serbyn13b,Ros15}, namely in MBL phase a
 complete set of local
 integrals of motions (LIOMS) may be defined. On one side finding LIOMs provides 
 information about the system for a given disorder realization (LIOMs are disorder realization dependent) -- on 
 the other side the very existence of LIOMs explains the Poissonian statistics observed deep in the localized phase.
  While the two extremal situations -- the metallic, GOE-like ergodic behavior for a weak disorder and the full 
 MBL phase seem to be presently quite well understood it is desirable to understand and describe the
 nature of the ergodic-MBL transition. 
 
 The problem is not simple -- it has been found, in particular, that
 the nature of the disorder plays a decisive role in the character of the transition
 \cite{Khemani17,Zhang18}. Intra-sample randomness was indentified as the
 dominant feature for quasiperiodic disorder (QPD) while the inter-sample 
 randomness is an essential property of transition for purely random disorder (RD). Those important
 observations were made studying the entanglement entropy behavior.

 In this work, we show that a proper analysis of gap ratio statistics 
  allows us to  get similar insight on the randomness of system in MBL transitions as the 
  entanglement entropy \cite{Khemani17}. Our method is conceptually simpler
  as it relies only on the spectrum of the system and as such can be straightforwardly used in studies
  of various complex systems. Secondly, this analysis, as a by--product, 
  gives hints on the construction of  universal model of level statistics for MBL transition which we provide
   generalizing earlier attempts
 \cite{Shklovskii93,Serbyn16,Bertrand16}. 
 We introduce a weighted short--range plasma model (wSRPM) and argue that it 
 describes faithfully the level statistics during the whole 
crossover between ergodic and MBL phases at system sizes accessible in exact diagonalization studies.
Taking into account the inter-sample randomness -- an inherent feature of MBL transition
in systems with random disorder,
the proposed model
grasps correctly not only the bulk properties of 
the level spacing distribution $P(s)$ but also its exponential tails  and 
correctly reproduces the number variance, $\Sigma^2(L)$,
at $L$ of the order of tens level spacings. This implies that wSRPM reflects faithfully both short-range 
and long-range spectral correlations in systems across the ergodic-MBL crossover.
Remaining small discrepancies 
are discussed in details providing a further insight into the long range spectral correlations 
of the system. 
Furthermore, we show that the wSRPM is universal as it works across the whole ergodic to MBL crossover
 not only in spin models  but also
in disordered bosonic and fermionic systems.  
We compare our results with earlier propositions \cite{Chavda14,Serbyn16,Shukla16,Bertrand16,Buijsman18}
showing that the model proposed by us represents the data much more faithfully.
We also discuss an alternative model of level statistics -- weighted power-law random banded matrix model, which also 
accurately grasps spectral correlations across the ergodic-MBL crossover.

 \section{Gap ratio analysis}
 \label{sec: ranal}
 
A dimensionless ratio of consecutive energy levels
gaps  (referred as the gap ratio) was introduced in \cite{Oganesyan07}. It is 
defined as $r_n=\min\{\delta_n,\delta_{n-1}\} / \max\{\delta_n,\delta_{n-1}\}$ 
where  $\delta_n=E_{n+1}-E_n$ is an energy difference between two consequtive levels.
The average gap ratio, $\bar r$,  is different for systems with extended eigenstates
(in the following we shall concentrate on the gaussian orthogonal ensamble 
(GOE) for time-reversal invariant systems) $\bar r_{GOE} \approx 0.53$ and for 
localized systems $\bar r_{Poi} \approx0.39$ as was analyticaly demonstrated in \cite{Atas13}. That property
was used by many authors in attempts to localize the MBL transition
\cite{Oganesyan07,Pal10,Mondaini15,Luitz15,Luitz16,Sierant17,Sierant17b,Sierant18,Janarek18,Wiater18}.

 The usual way of calculating the mean gap ratio $\overline r$
is to average the $r_n$ variable over a certain number of energy levels getting a mean gap
ratio for one sample
$r_S = \langle r_n \rangle_{S}$. Then, the mean gap ratio is obtained by averaging of $r_S$ over
disorder realizations $\overline r = \langle r_S \rangle_{dis}$. While, as mentioned above
$\overline r$ obtained in this
way reflects the character of eigenstates of the system 
\cite{Oganesyan07,Pal10,Mondaini15,Luitz15,Luitz16,Sierant17,Sierant18} a
part of information encoded in the $r_n$ variables is necessarily lost.
{Let us examine} $P(r_S)$ -- the distribution of the sample averaged gap ratio $r_S$ -- it provides a 
direct information {about variations 
of the $r_S$ for different disorder realizations}. As an example {we} consider  
the XXZ spin-1/2 chain with 
additional next-nearest-neighbors coupling (similar to that of  \cite{Khemani17})
\begin{equation}
 H= J\sum_{i=1}^{K} \vec{S}_i \cdot \vec{S}_{i+1} + W\sum_{i=1}^{K} \cos(2\pi \zeta i + \phi) S^z_i+ J_1\sum_{i=1}^K S^z_i S^z_{i+2},
 \label{kheH}
\end{equation}
{where}  $\vec{S}_i$ are spin-1/2 matrices,  $\zeta= (\sqrt{5}-1)/2$ (the golden ratio) and $\phi$ is
a fixed phase for a given disorder realization (leading to QPD)
or is random on each lattice site (leading to RD with 
the same on--site distribution, as in 
the QPD case) {\cite{Khemani17}}. We fix $J=1$ as the energy unit and 
we study the case of
$J_1=J$ first. Periodic boundary conditions 
are assumed so that $\vec{S}_{K+1} = \vec{S}_1$.
 \begin{figure}
\includegraphics[width=1\columnwidth]{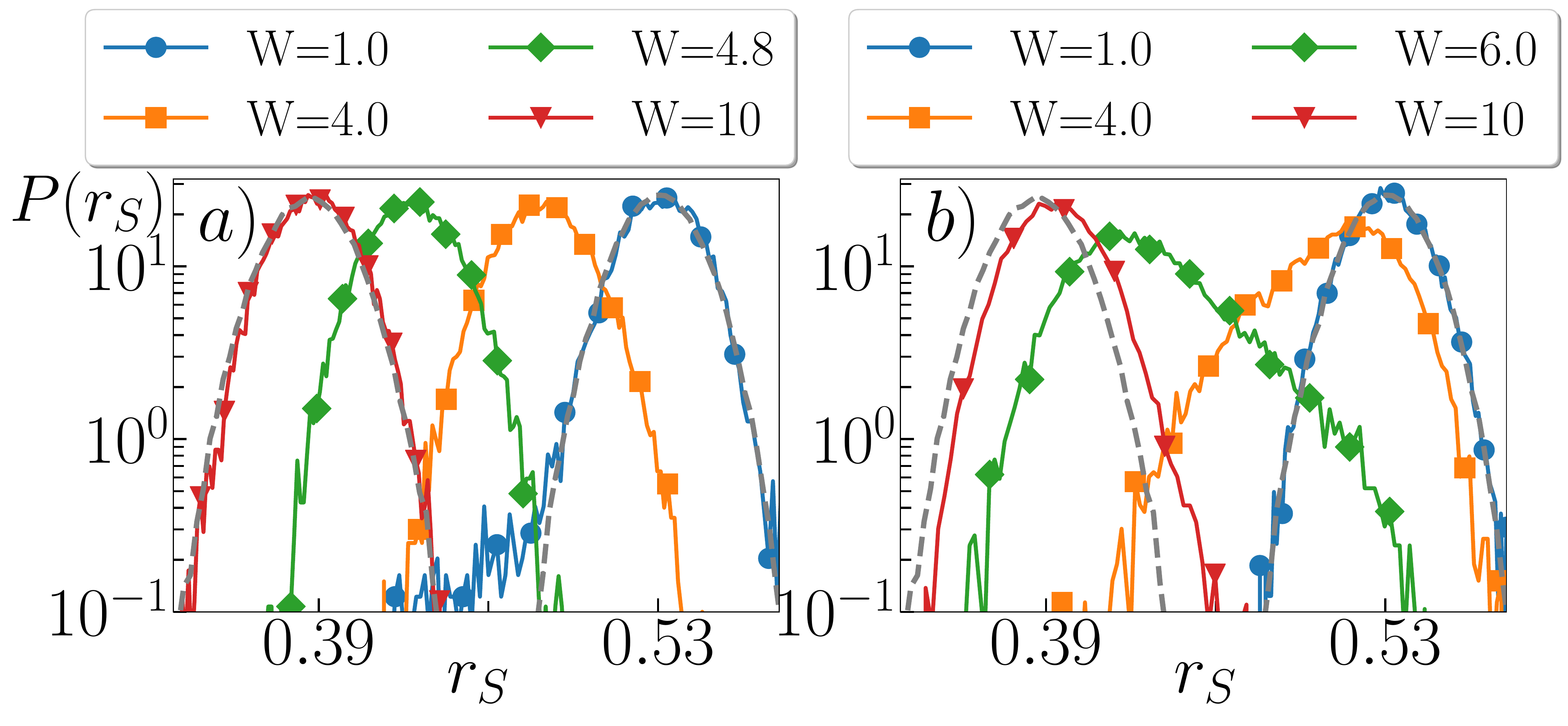}
\includegraphics[width=0.5\columnwidth]{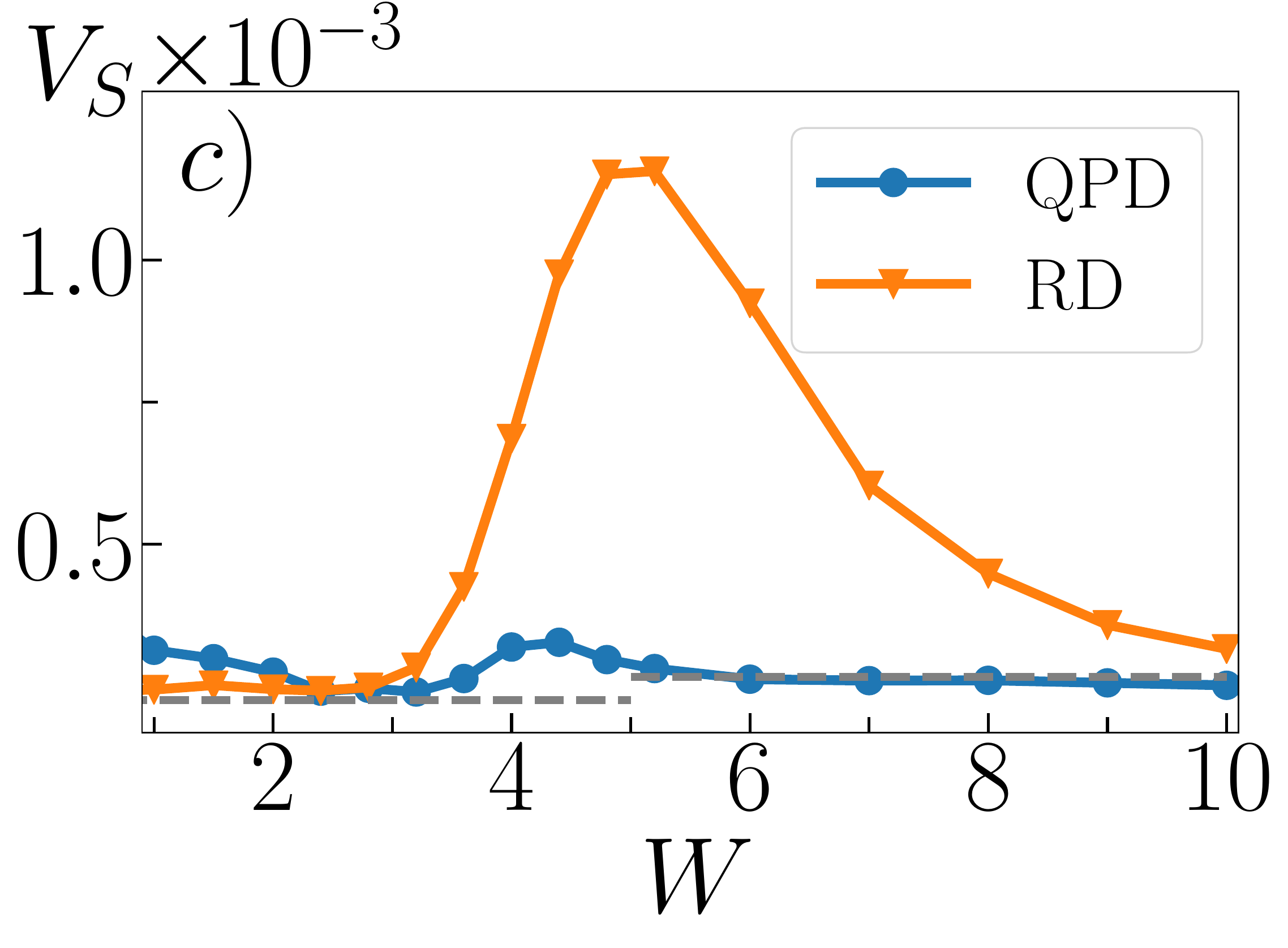}\includegraphics[width=0.5\columnwidth]{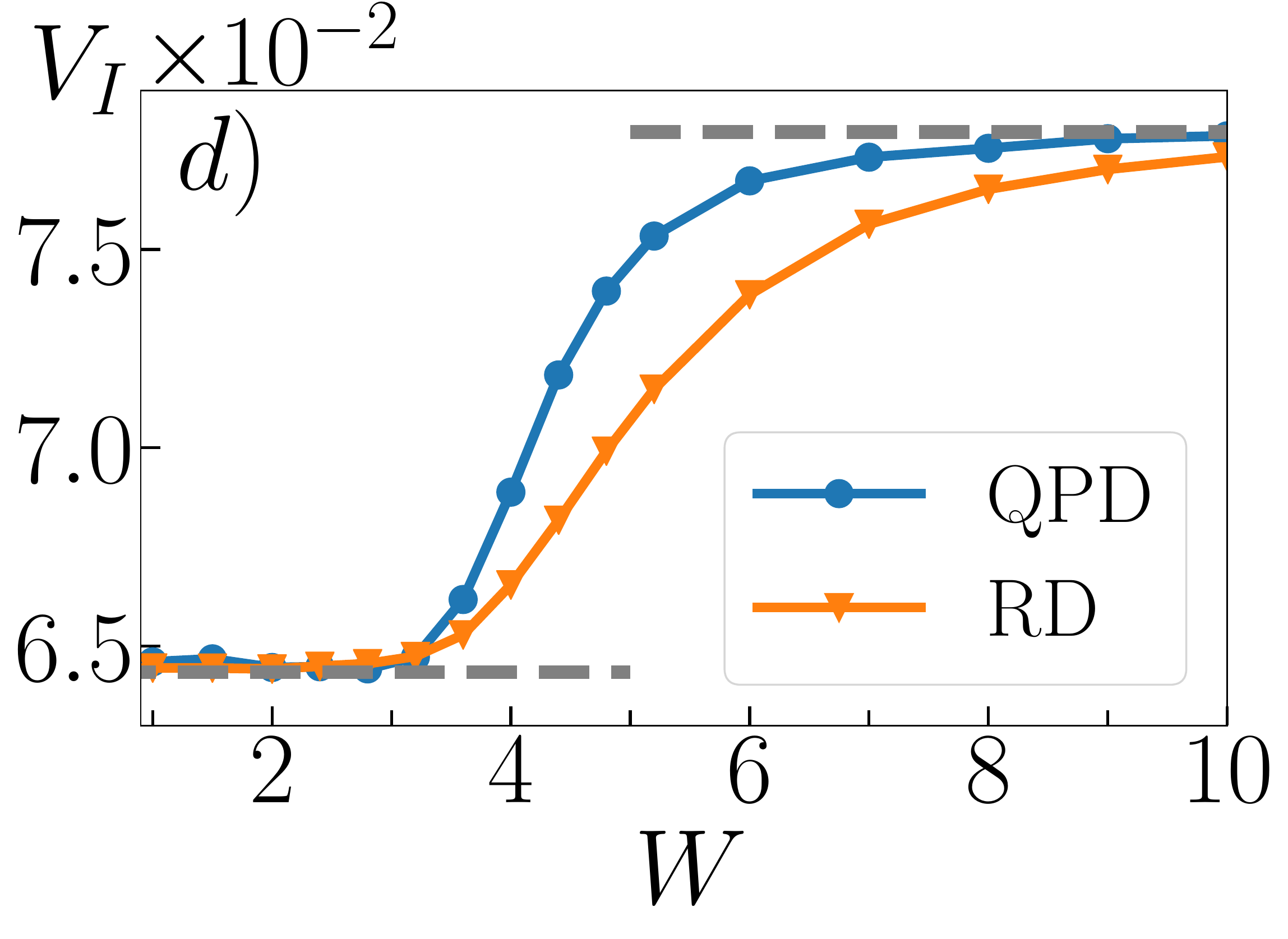}
\caption{ \label{fig: QR1} Top: (a) the distributions $P(r_s)$ for quasi-periodic disorder (QPD) of
strength $W$. Dashed lines give limiting GOE and 
Poisson  behaviors. The tail of the $W=1$
distribution indicates that QPD reproduces GOE statistics only approximately; 
(b) the distributions $P(r_s)$ for random disorder (RD).  Bottom: (c) the inter-sample variance $V_S$
for RD and QPD; (d)
  the intra-sample variance $V_I$  for QPD and RD during the transition.
}
\end{figure} 
For the system size $K=16$ we consider {sequences of} $N=400$ consecutive 
eigenvalues from the middle of the spectrum yielding a {collection} of $r_S$ values for  {$n_{dis}=2000$} disorder realizations. 
The resulting distributions,
 $P(r_S)$, for different disorder strengths $W$
 are shown in  Fig.~\ref{fig: QR1}.

 Had all $r_n$ been independent of each other the distribution of $r_S=\sum_{n=1}^N r_n/N$ should be Gaussian
 with width determined by the variance of the $r_n$ distribution and proportional to $1/\sqrt{N}$. 
Despite the correlations -- particularly strong for GOE -- the $P(r_S)$ are Gaussian in 
the limiting cases of GOE and Poisson statistics. Surprisingly, the $P(r_S)$ distributions  
remain Gaussian for QPD across the transition.

In a striking contrast, 
the distributions {in the RD case} become strongly asymmetric with  enlarged variance in the transition region.
This reflects the inter{-}sample randomness importance for the RD and is a clear, nice manifestation of
the existence of rare Griffiths regions \cite{Griffiths69,Vojta10,Agarwal15,Agarwal16}: for samples
with $\bar r$ close to GOE there exist realizations of disorder leading to $r_S$ close to Poisson limit.
Similarly, on a localized side for  $\bar r$ close to integrable limit there are rare events with $r_S$
values close to  GOE value. 
{The stark difference in the $P(r_S)$ distributions between the RD and QPD cases can be quantified by 
calculating a variance: $V_S = \langle r_S^2 - \overline r ^2 \rangle_{dis}$. As Fig.~\ref{fig: QR1}c) shows, the inter-sample 
variance $V_S$ has a clear peak in the MBL transition for the RD whereas it varies only slightly for the QPD.}

Consider now  the variance $v_I$ 
of the $r_S$ variable, $v_I=\langle r_n^2-r_S^2 \rangle_{S}$. Averaged over disorder realizations
$V_I = \langle v_I \rangle_{dis}$, it provides information about fluctuations of $r_n$
within {a} single spectrum of the system at a certain disorder strength -- characterizing intra-sample randomness.
As could be expected from the 
long range correlations of GOE, it is small for GOE and conversely, it is maximal for Poissonian spectrum. 
Fig.~\ref{fig: QR1}{d) shows that} it behaves
similarly for QPD and RD {interpolating between the values for GOE and Poisson statistics.}
The transition is sharper for the system with QPD, implying that it is less affected by 
finite size effects \cite{Khemani17}. 

\begin{figure} 
 \includegraphics[width=0.5\columnwidth]{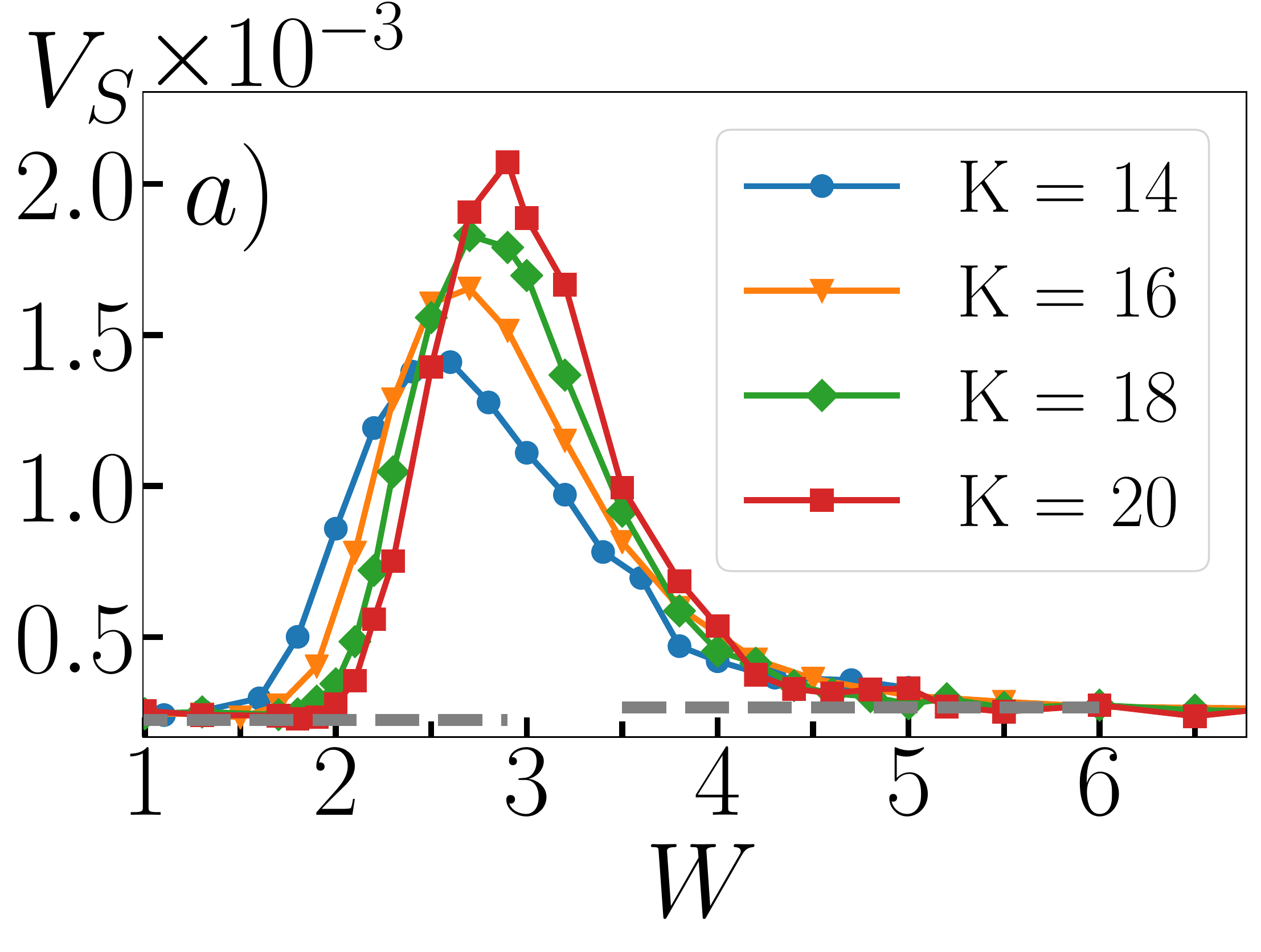}\includegraphics[width=0.5\columnwidth]{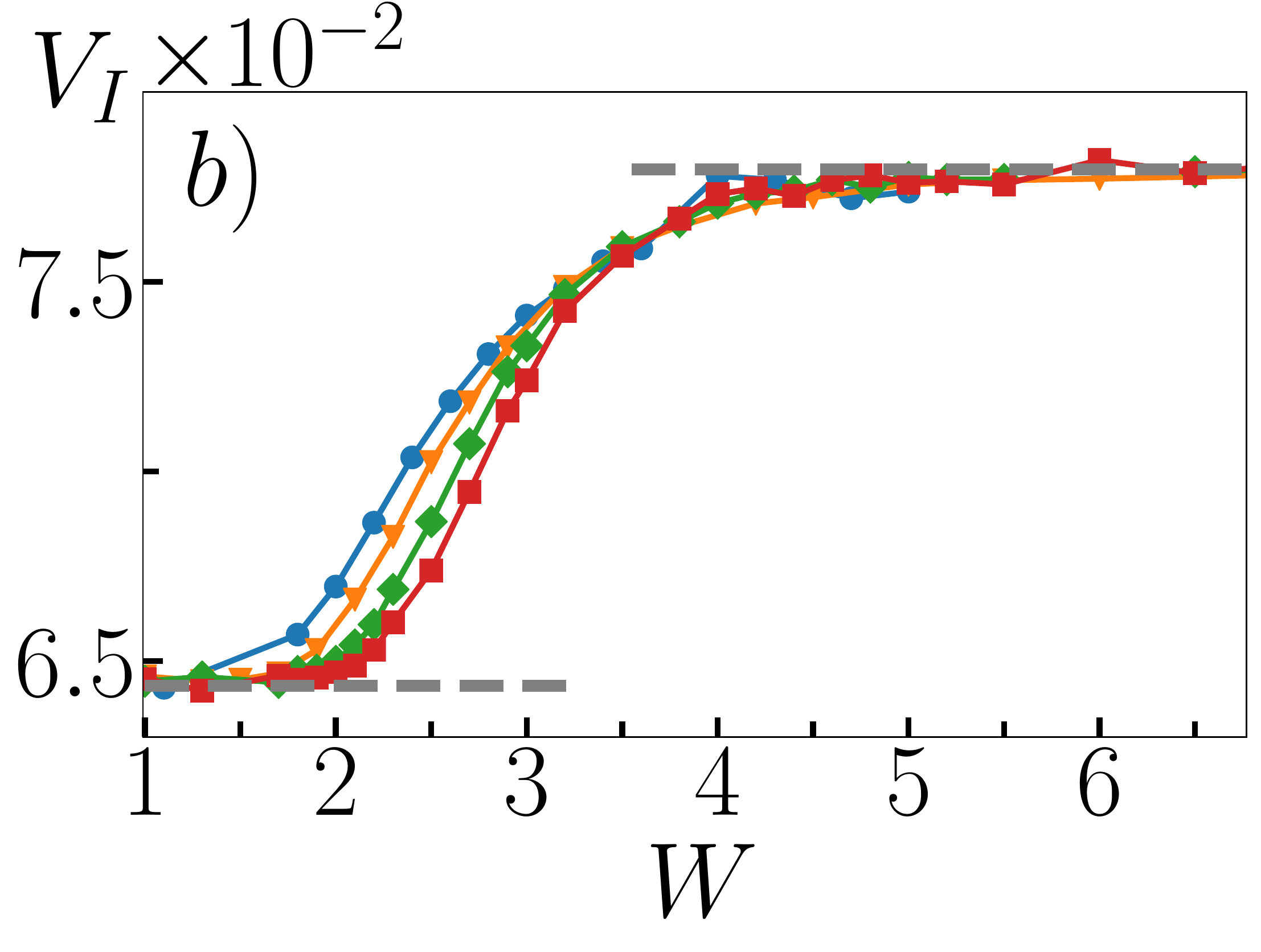}
\includegraphics[width=0.5\columnwidth]{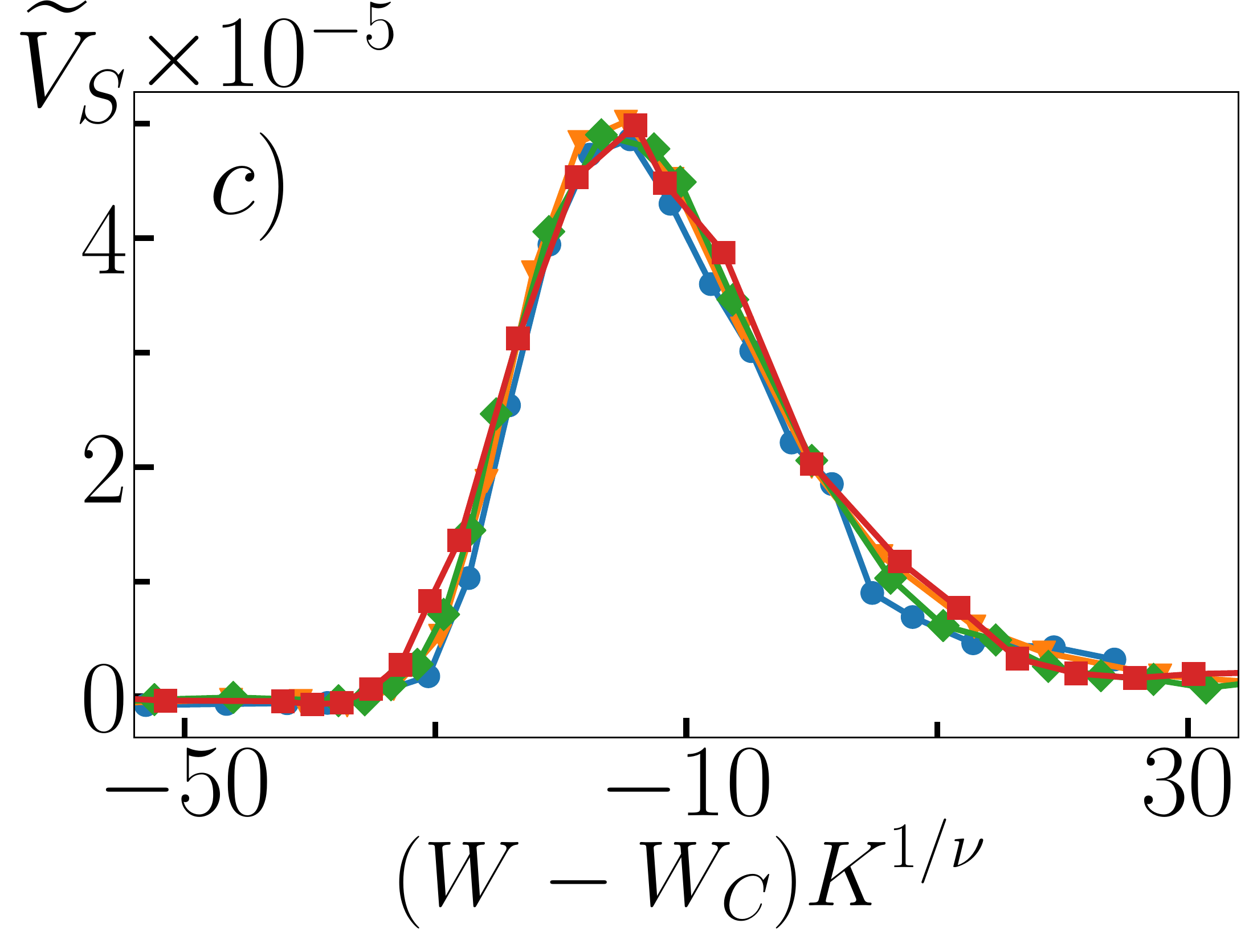}\includegraphics[width=0.5\columnwidth]{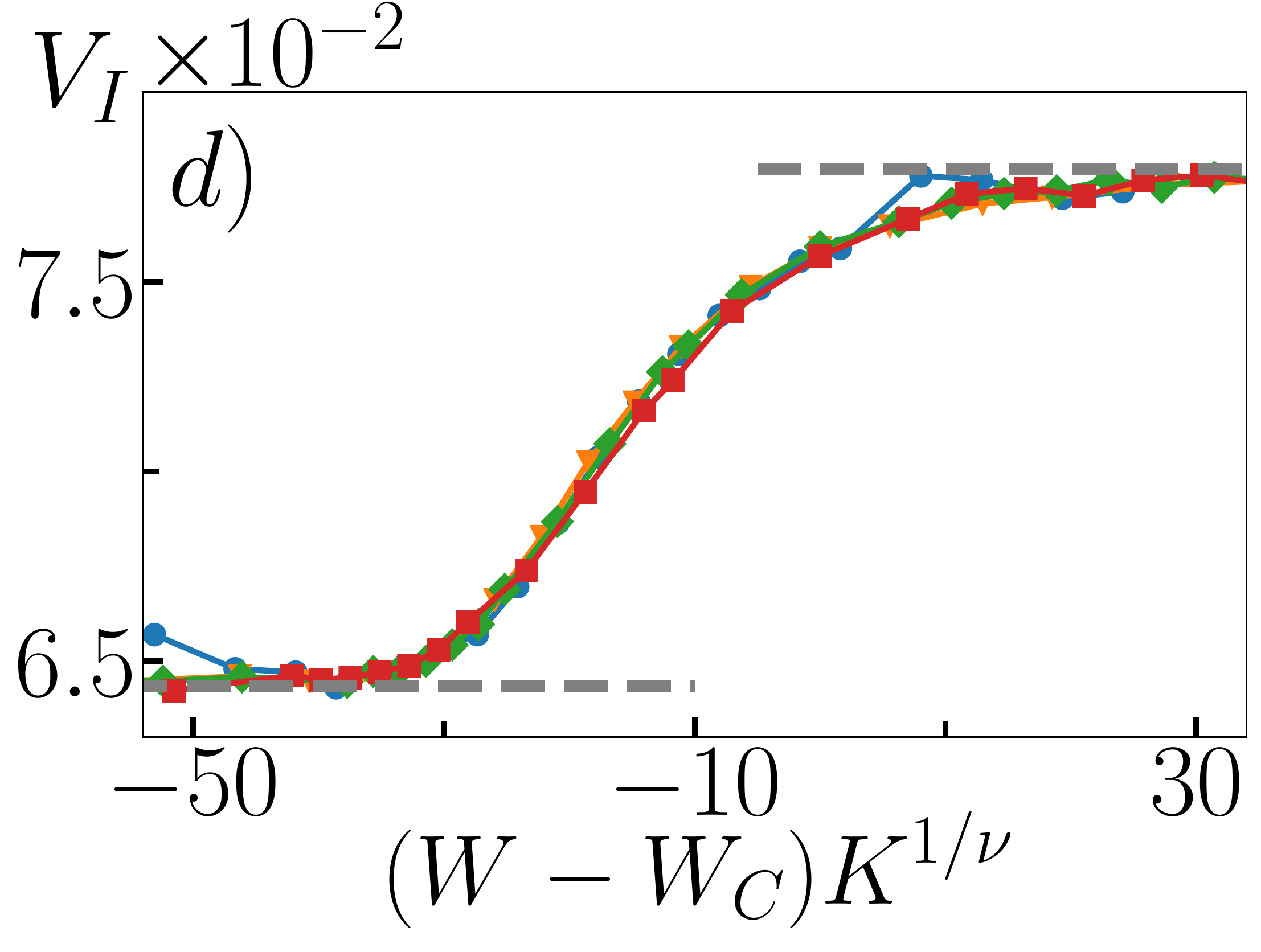}
\caption{\label{fig: Var1} {Top: (a) The  variance $V_S$ of the $r_S$ distribution 
characterizing the inter-sample randomness; (b) the variance $V_I$ reflecting the intra-sample
fluctuations in the spectrum of the system.
 Bottom: (c) the rescaled inter-sample variance $\widetilde{V}_S$ and (d) the intra-sample variance $V_I$
 collapse after the rescaling of the disorder
strength with $W_C = 3.5$ and $\nu = 0.95$.
 Data for system sizes $K\in\{14,16,18,20\}$. } }
\end{figure}

Seeing that the distribution $P(r_S)$ and the variances $V_S$ and $V_I$ provide a valuable information 
about the randomness at the {MBL} transition, let us switch our attention to the more
standard Heisenberg chain case taking $J_1=0$ in Eq.~\eqref{kheH} and assuming 
random uniform disorder so that $W\cos(2\pi \zeta i + \phi)$ is exchanged by $h_i \in [-W,W]$ in Eq.~\eqref{kheH}, explicitly
\begin{equation}
 H= J\sum_{i=1}^{K} \ \vec{S}_i \cdot \vec{S}_{i+1} + \sum_{i=1}^{K} h_i S^z_i.
 \label{eq: XXZ}
\end{equation}
Despite the fact that the distribution of disorder is different and the studied model  contains now nearest 
neighbor couplings only, the $P(r_S)$ behaves quite similarly to the case shown in 
Fig.~\ref{fig: QR1}b) revealing strong asymmetry and broadening across the transition -- as shown in
Fig.~\ref{fig: Var1}. Particularly, the 
broader distributions in the transition regime suggest that one may use the maximal variance $V_S$ as an 
indicator of the transition point.

A standard  finite size scaling of different quantities can be performed assuming $W \rightarrow (W-W_C)K^{1/\nu}$. 
For $\bar r$ such an analysis has been performed 
already \cite{Luitz15,Kudo18} with the data collapsing to a single curve.
Similar scaling may be used for  the variance $V_S$. Observe that both the position of
the maximum as well as its value depend on the system size -- Fig.~\ref{fig: Var1}a).
{If, together with 
the rescaling of the disorder strength, 
the variance $V_S$ is rescaled according to $V_S \rightarrow \widetilde{V}_S=(V_S-V_{GOE})/K^{\kappa}$
(where $V_{GOE}$ is the inter-sample variance for GOE) the data for various system sizes collapse onto a single 
curve -- Fig.~\ref{fig: Var1}c) for the exponents $\nu=0.95(10)$, $\kappa = 1.2(1)$ and the critical disorder
strength $W_C=3.5(1)$. The scaling of the $V_S$ will necessarily cease to work for larger system sizes as the support of 
the $P(r_S)$ distribution is limited by $\overline r_{Poi}$ and $\overline r_{GOE}$. On the other hand, the 
critical disorder strength $W_C=3.5(1)$ and the exponent $\nu=0.95(10)$ are in nice agreement with results 
of \cite{Luitz15}.}
A similar finite size scaling may be performed 
for {the} intra-sample variance $V_I$ with the same $W_C$ and $\nu$ -- Fig.~\ref{fig: Var1}d). 
It is notable that all three measures $\overline r$, $V_S$ and $V_I$ scale in a very similar manner. 
Being interconnected they still provide different insights into physics of the system during the MBL transition.  

The gap ratio analysis demonstrates that more than just 
 an overall information about the crossover between ergodic and MBL regimes can be obtained from the 
 $r_n$ variables. The considered inter- and intra-sample variances $V_S$ and $V_I$ 
 reflect nicely the differences between RD and QPD universality classes. Furthermore, the $P(r_S)$ distribution
 quantifies the inter-sample fluctuations of a system undergoing MBL transition and gives a particularly 
 clear demonstration of the Griffiths regime. 
 
 Moreover, the gap ratio analysis hints how to formulate 
 the wSRPM model of spectral statistics across the MBL transition for the random disorder -- namely, the 
 ensemble we are looking for should take into account the large inter-sample randomness of the RD case.
 The problem of construction of such an ensemble  will be considered in the remaining part of the manuscript. 
 We start by reviewing the existing models of level statistics in the MBL transition.

 \section{Level statistics in MBL transition}

\begin{figure}
\includegraphics[width=1\columnwidth]{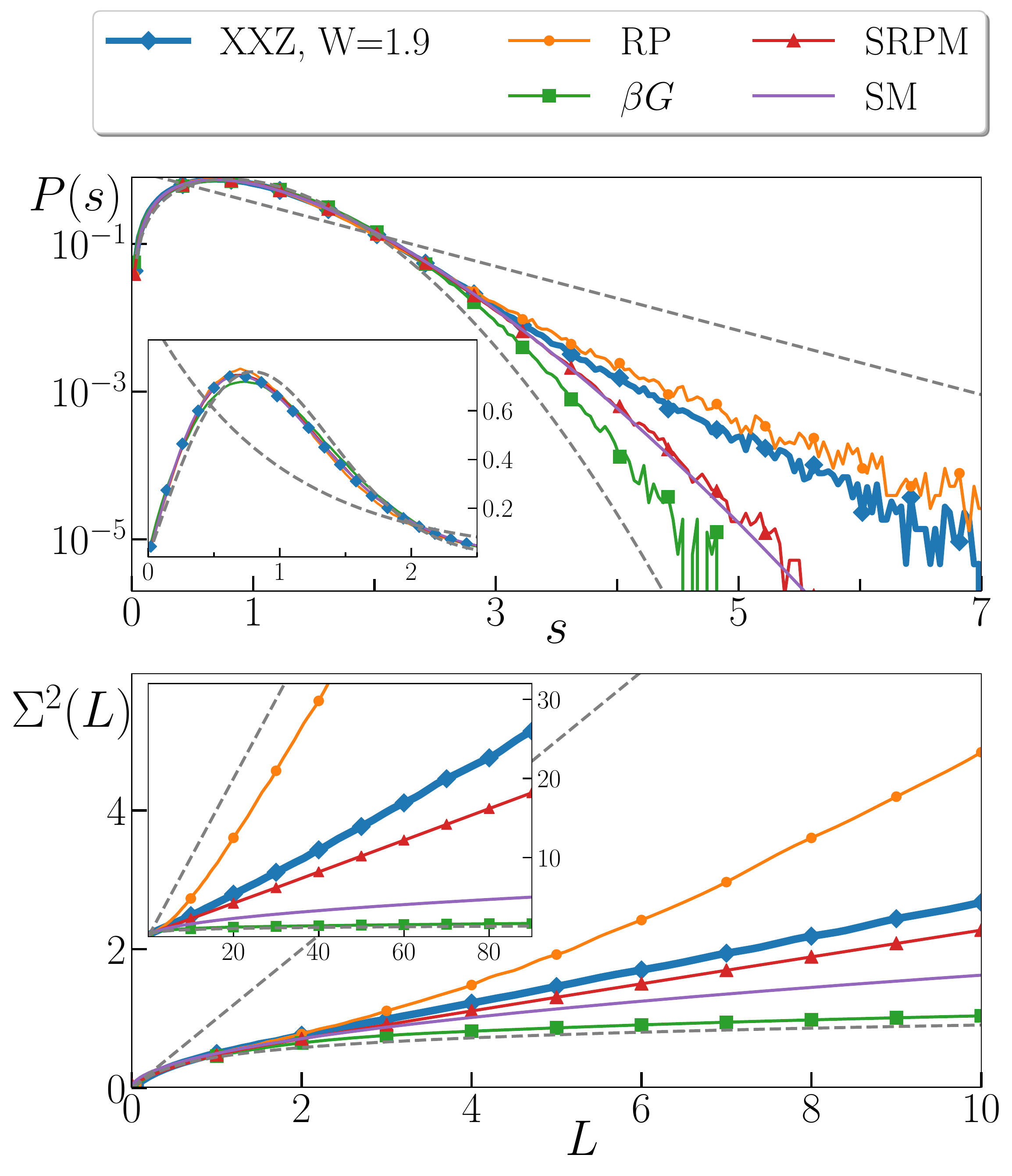}
\caption{ \label{fig: cmp} Top: level spacing distribution $P(s)$ for XXZ spin chain \eqref{eq: XXZ} of size $K=16$ 
at disorder strength $W=1.9$ compared with predictions of various models of 
{the flow of level statistics between GOE and PS limits in the MBL crossover} discussed in the main text. The 
vertical axis of the main plot is logarithmic to enable the comparison of tails of the distributions, inset 
shows the data in doubly linear scale; RP --
the Rosenzweig-Porter model at $\sigma=0.0016$; $\beta G$ -- $\beta$ Gaussian ensemble with $\beta =0.81$;
SRPM -- short range plasma model with the range of interactions $h=5$; SM -- the mean field model \eqref{eq: SM1}.
Bottom: the number variance $\Sigma^2(L)$ for the same system, inset shows the long range behavior of $\Sigma^2(L)$
which encodes long range spectral correlations of eigenvalues. Gray dashed lines correspond to level spacing 
distributions and number variances for GOE and PS limits. 
}
\end{figure}

 A number of models for intermediate statistics in the MBL transition have been proposed in
\cite{Serbyn16,Bertrand16,Shukla16,Buijsman18}. In this section we compare level spacing distributions
$P(s)$ and number variances $\Sigma^2(L)$ predicted by those models with data for the standard
model of MBL -- XXZ spin-1/2 chain  Eq.~\eqref{eq: XXZ}, studied already in the previous section. As we have seen this model undergoes a transition to MBL
at $W_C \approx 3.5$ in the thermodynamic limit
$K\rightarrow \infty$, in the center of the spectrum ($W_C\approx 3.7$ was obtained in \cite{Luitz15}). Fig.~\ref{fig: cmp}
shows level spacing distribution $P(s)$ and number variance $\Sigma^2(L)$ for the system \eqref{eq: XXZ} of size $K=16$
at disorder strength $W=1.9$ compared with predictions of different proposed models \cite{Serbyn16,Shukla16,Buijsman18} 
of the flow of level statistics between GOE and PS limits
supplemented by data for the short-range plasma model (SRPM) \cite{Bogomolny01}.
The numerical data for spacing distribution and the number variance for XXZ spin chain 
are fitted with those models.

\noindent{\it Mean field plasma model. }
The work \cite{Serbyn16} describes the flow of level statistics across the MBL transition.
 Close to the ergodic regime
a mean field plasma model \cite{Kravtsov94} with an effective power--law interaction between energy levels
is proposed. It predicts  the level spacing distribution and the number variance to be
\begin{equation}
P(s) = C_1 s^{\beta} \mathrm{e}^{-C_2 s^{2-\gamma}} \,\,\,\,\, \mathrm{and} \,\,\,\,\,\, \Sigma_2(L) \propto L^{\gamma}
 \label{eq: SM1}
\end{equation}
with $C_{1,2}$ determined by normalization conditions $\langle 1 \rangle=\langle s \rangle = 1$.
The exponents $\beta$, $\gamma$ reflect a local repulsion of energy levels and an effective range
of interactions between energy levels. They are treated as fitting parameters which vary across the transition.  Note that for $\gamma=1$ the eigenvalues are interacting only locally
leading to semi--Poisson statistics
\begin{equation}
P(s) \propto
s^{\beta } \mathrm{e}^{-(\beta +1) s}
\,\,\,\,\, \mathrm{and} \,\,\,\,\,\, \Sigma_2(L) \propto \frac{1}{\beta+1}L.
 \label{eq: SM2}
\end{equation}
Between GOE and Poisson limits the exponent $\gamma$ satisfies 
 $ 0 < \gamma < 1$. {It follows from} \eqref{eq: SM1}  that tail of the level spacing distribution 
decays faster than exponentially with $s$ and  that the number variance $\Sigma^2(L)$ increases as a power law of $L$.
The level spacing distribution and the number variance predicted by this model 
 are denoted by the solid violet line in Fig.~\ref{fig: cmp} -- the values of $\beta$ and $\gamma$ are obtained by the least square fit 
to the bulk of $P(s)$ and the multiplicative factor in front of the $\Sigma^2(L)$ is treated as the third 
fitting parameter. While the bulk of the level 
spacing distribution is nicely recovered, the tail of the $P(s)$ distribution 
and the number variance are clearly not matching the data for $W=1.9$.

This two features were shown to be not obeyed by a system {across the MBL crossover} 
 in \cite{Bertrand16} where it was demonstrated
that the level spacing distributions decay exponentially with $s$. {At the same time}  the number variance  
increases as $L^{\gamma}$ with $\gamma > 1$ close to the ergodic phase and as the system becomes more localized it
{becomes}
asymptotically linear for large $L$.

\noindent{\it Rosenzweig-Porter ensemble. } Another work \cite{Shukla16} suggest
that Rosenzweig-Porter (RP) ensemble can be appropriate to describe 
the MBL transition. Multifractal properties of eigenvectors of this model, which 
is defined as an ensemble of  real symmetric  (for $\beta=1$ orthogonal class relevant for us)
 random 
matrices $M=\left( M_{ij} \right)$ of size $n\times n$ with matrix elements being  independent Gaussian variables 
with zero average values $M_{ij}=0$ and 
\begin{equation}                                                                                              
\langle M_{ii} ^2 \rangle = 1, \,\,\,\,\,\,\,\, \mathrm {and} \,\,\,\,\,\,\,\,\,                  
\langle M_{ij}^2 \rangle =\sigma/2
\label{eq: RP}                                                                                             
\end{equation}      
were studied in \cite{Kravtsov15}.
The dotted line in Fig.~\ref{fig: cmp} shows the obtained level 
spacing distribution and the number variance  
which fits best the data for the XXZ spin chain at $W=1.9$. The presented data are for $n=3000$
 and $\sigma=0.0016$
and are in rather poor agreement even regarding the bulk of the $P(s)$ distributions. 
Moreover, at $ L\gtrsim3$ the number variance 
bends abruptly upwards -- a feature which we do not observe for the $W=1.9$ data.
{Similar trend persists at larger disorder strengths $W$ indicating that one cannot reproduce 
 both the 
 level spacing distribution and the number variance $\Sigma^2(L)$ of the $XXZ$ spin chain across the MBL crossover
within the RP ensemble.}

\noindent{\it $\beta$-Gaussian ensemble. }
The two remaining models \cite{Bogomolny01,Buijsman18} can be specified by a joint probability 
distribution function (JPDF) of eigenvalues. 
A JPDF for a Random Matrix Ensemble can be written as 
the   probability   distribution   of   a   one-dimensional  gas  of  classical  particles  
with total energy $W(E_1,..., E_n)$
\begin{equation}
 \mathcal P (E_1,..., E_N) = Z_N^{-1}  \exp\left( - \beta W(E_1,...,E_n) \right),
 \label{eq: JPDF1}
\end{equation}
where $Z_N$ is a normalization constant and the total energy
\begin{equation}
 W(E_1,...,E_n) = \sum_i U(E_i) + \sum_{i<j}  V(|E_i-E_j|)
 \label{eq: JPDF1a}
\end{equation}
is determined by the trapping potential $U(E)$ and inter-particle interactions $V(|E-E'|)$. 
For instance, for harmonic trapping potential
$U(E) \propto E^2$, and logarithmic interactions  $V(|E-E'|) = -\log(|E-E'|)$ and $\beta=1$ one
recovers {from} \eqref{eq: JPDF1} the 
JPDFs for GOE, for which the interactions in \eqref{eq: JPDF1} are between all pairs of eigenvalues which reflects
the long range spectral correlations of the GOE ensemble.

One way of constructing an ensemble with statistical properties intermediate between GOE and PS is
to put a rational $\beta \in[0,1]$ into JPDF \eqref{eq: JPDF1} -- in such a way a $\beta$-Gaussian ensemble ($\beta$GE) 
arises. 
A recent work \cite{Buijsman18} uses $\beta$GE  to describe the {level spacing distribution}
$P(s)$ and the gap ratio 
distribution in the MBL transition. Setting up appropriate tridiagonal matrices \cite{Dumitriu02} 
of size $n=10^5$ and diagonalizing them, we obtain $P(s)$ and $\Sigma^2(L)$ for this ensemble --
denoted by the green line with squares in Fig.~\ref{fig: cmp}. The agreement 
of this model with XXZ numerical data in the bulk of the $P(s)$ is
not perfect. The disagreement in the tail of the $P(s)$ and the number variance is even more pronounced.
Long--range correlations of eigenvalues in $\beta$GE are visible in the spectral rigidity of
the spectrum -- for the acquired data the number variance grows only logarithmically, 
just like in the GOE case, in a violent disagreement with the XXZ data. Thus, 
contrary to statements in \cite{Buijsman18} based on short range correlations only, the $\beta$GE  
is not a good candidate to describe the flow of level statistics between GOE and PS regimes
across the MBL transition.

\noindent{\it Short-range plasma models. }
Another way of constructing intermediate level statistics is to restrict the range
of the logarithmic interactions in \eqref{eq: JPDF1}
to a finite number $h$ which leads to a family of short-range plasma models (SRPMs) \cite{Bogomolny01}.
Consider $N \rightarrow \infty$ particles in a ring geometry
$E_0 < E_1 < ... < E_N < E_{N+1}$, $\,E_{N+1+k}=E_{k} \mod N$ 
with logarithmic interaction among $h$ neighboring eigenvalues so that the JPDF is 
given by 
\begin{equation}
 \mathcal P_h^{\beta} (E_1,..., E_N) = Z_N^{-1}  \prod_{i=0}^{N} |E_i - E_{i+1}|^{\beta} ... |E_i - E_{i+h}|^{\beta}.
 \label{eq: SRPM}
\end{equation}
For integer values of $h$ and $\beta$ this model can be analytically solved yielding the level 
spacing distribution
\begin{equation}
 P_h^{\beta}( s ) = s^{\beta} W(s) \mathrm{e}^{-(h\beta+1)s}
 \label{eq: PS1}
\end{equation}
where $W(s)$ is a polynomial. The corresponding number variance has asymptotically linear behavior: 
\begin{equation}
 \Sigma^2_{h,\beta}(L) \stackrel{L\rightarrow \infty}{\longrightarrow} \frac{L}{h\beta+1}.
 \label{eq: PS2}
\end{equation}
{The SRPM can be solved analytically -- see  Appendix \ref{AppA} for details.}
While grasping
the bulk of {the level spacing distribution} $P(s)$ accurately, the SRPM model does not
outperform the mean field model (\ref{eq: SM1}-\ref{eq: SM2}). 
One still does not obtain the correct tails of the level spacing {distribution} $P(s)$ or
the correct slope of the number variance $\Sigma^2(L)$ -- see the line with triangles in Fig.~\ref{fig: cmp}.

\section{The weighted SRPM model}

The preceding section shows that the analyzed models reproduce the bulk of the 
level spacing distribution and hence grasp purely local correlations of
eigenvalues of a system in ergodic to MBL crossover. However, when tails of the level spacing distributions
as well as the number variance are considered, the differences between data for XXZ spin chain
and the predictions of the models are apparent. This shows that the models do not faithfully reproduce correlations
between eigenvalues on scales larger than few level spacings. The mean field plasma model, $\beta$-Gaussian ensemble
and SRPMs tend to underestimate the number variance predicting stronger long-range correlations between eigenvalues than
are actually observed across the MBL crossover. The opposite is true for the RP ensemble. All in all, the SRPM with its asymptotically 
exponentially decaying level spacing distribution and asymptotically linear number variance gives predictions 
closest to the data for XXZ spin chain. For that reason we choose the SRPM as a basic building block of a 
more complicated, \textit{weighted}
ensemble which, by construction, takes into account another feature of the MBL crossover -- the large inter-sample randomness. 

Results of Sec.~\ref{sec: ranal} indicate that large inter-sample randomness is an inherent 
feature of the MBL transition in systems with purely random disorder. It manifests itself in 
shape of a distribution $P(r_S)$ of the gap ratio  for a single disorder realization $r_S =\langle r_n \rangle_S$ 
 which
significantly broadens in the regime of MBL transition. 
The broadening of $P(r_S)$ shows that system which has 
predominantly ergodic features becomes more localized for certain disorder realizations -- the 
converse statement for mostly localized system is also true.
The small fraction of events for which the system is more localized than usually reveals itself
in the tail of the level spacing distribution and in the number variance.
For instance consider an ensemble of matrices created in such a way that with probability $1-p$
the matrix is taken from GOE and with probability $ p \ll 1 $ it has the Semi-Poisson level statistics $\mathcal 
P^{\beta=1}_{h=1}$. The bulk of the level spacing distribution of such an ensemble will 
be very close to the Wigner distribution $P_{GOE}(s)$ of the GOE matrix ensemble (as $p\ll 1$). However, 
 for large level spacings the distribution 
will be dominated by  exponentially decaying tail of the 
level spacing distribution $P(s)$ from the small fraction of matrices with Semi-Poisson 
statistics. Analogously, the number variance $\Sigma^2(L)$ will be a sum of logarithmically growing number variance
for GOE and linearly increasing number variance for Semi-Poisson statistics. Hence, it will be dominated by the latter
and increase linearly with $L$ with a very good approximation.

\begin{figure}
\includegraphics[width=1\columnwidth]{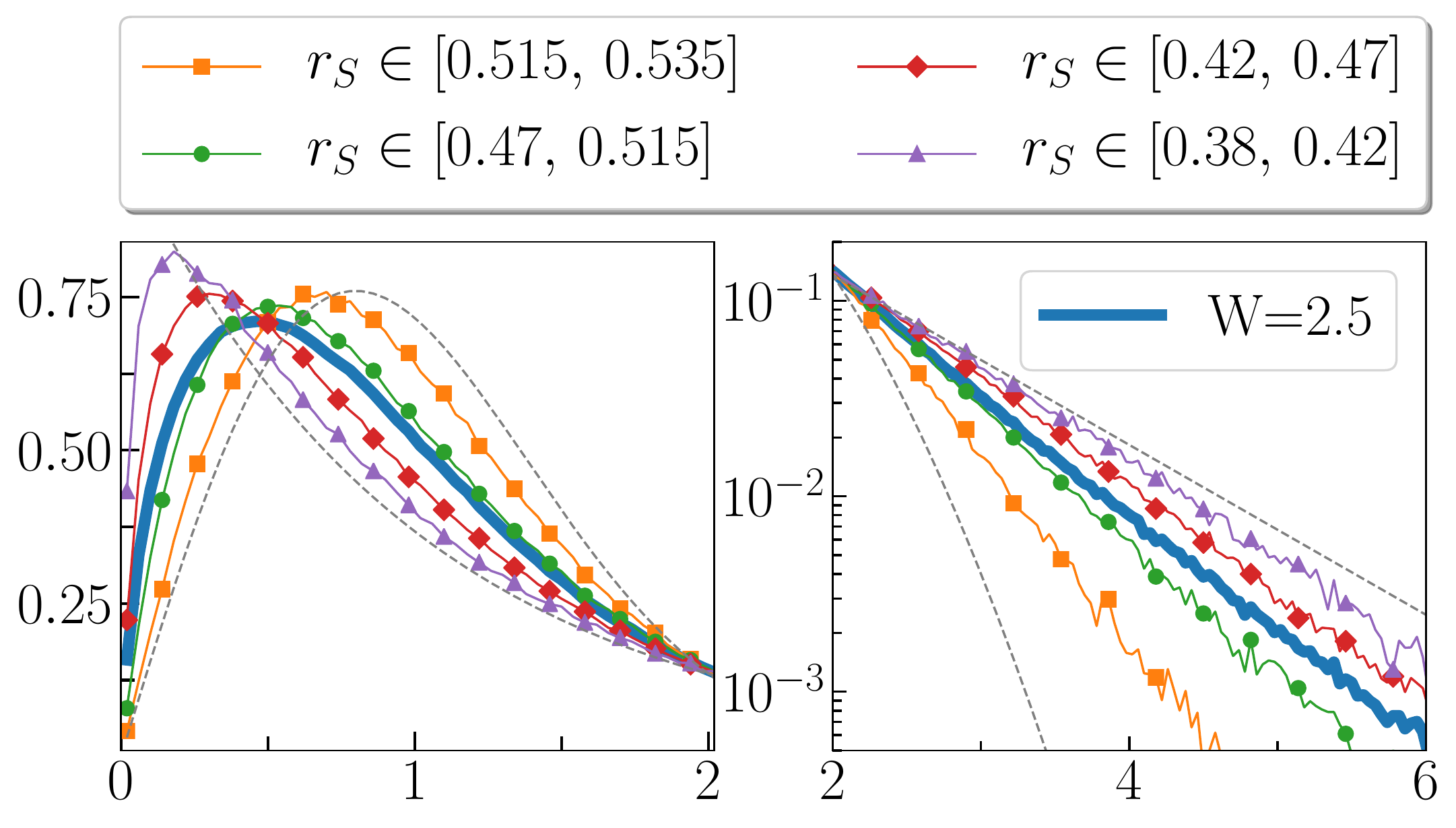}
\caption{ \label{fig: rfilter} Level spacing distribution $P(s)$ for XXZ spin chain \eqref{eq: XXZ} of size $K=16$ 
at disorder strength $W=2.5$ (solid blue line) -- left: lin-lin scale, right: lin-log scale to facilitate
comparison of tails of $P(s)$. Selecting disorder realizations for which $r_S$ is from a given interval
results in statistics with properties which vary between those of an ergodic and nearly localized system. 
{The gray dashed lines correspond to level spacing distributions of GOE and PS.}
}
\end{figure}

This leads us to a question whether the inter-sample randomness 
 can be responsible for the exponential tails of level spacing distribution and a linear number variance
in the MBL transition via the mechanism described above.
To verify this, we examine level statistics of XXZ system at certain disorder strength 
but accept only disorder realizations for which the $r_S$ belongs to a certain narrow interval -- results for $W=2.5$
are presented in Fig.~\ref{fig: rfilter}.
The procedure of selecting $r_S$ affects significantly the resulting level statistics. 
As the interval of $r_S$ shifts towards smaller
values of $r_S$ one obtains level spacing distributions 
with weaker and weaker level repulsion characterized by decreasing $\beta$ and with 
growing weight of the exponential tail.
Both features are precisely the asymptotic characteristics of the level spacing distribution for SRPM
\eqref{eq: PS1} for appropriately chosen $\beta$ and $h$.
However, our goal is to reproduce the full level statistics. 
An appropriate model should thus combine the contributions from disorder realizations with different
localization properties reflected by the varying value of $r_S$. 

This leads us to the formulation 
of the weighed short-range plasma model (wSRPM)  which, by definition, has JPDF given by
\begin{equation}
  \mathcal P_{gSRMP}(E_1,.., E_N) = \sum_i c_i \mathcal P_{h_i}^{\beta_i}(E_1,.., E_N)
  \label{eq: wSRPM}
 \end{equation}
 where $h_i$ and $\beta_i$ range over an appropriate set of values and $c_i$ are weight coefficients ($\sum_i c_i=1$). 
 {The weight coefficients are determined by the requirement that the wSRPM reproduces the inter-sample
 randomness reflected by the $P(r_S)$ distribution.}
 By integrating the JPDF for wSRPM with $\delta(s-|E_k-E_{k-1}|)$ one gets the level spacing 
 distribution
 \begin{equation}
  P_{wSRPM}(s) = \sum_i c_i P_{h_i}^{\beta_i}(s)
  \label{eq: wSRPM_P}
 \end{equation}
 which is a linear combination of the level spacing distributions $ P_{h_i}^{\beta_i}(s)$. 
 An analogous expression holds for the number variance 
\begin{equation}
  \Sigma^2_{wSRPM}(L) = \sum_i c_i \Sigma^2_{\beta_i,h_i}(L),
  \label{eq: wSRPM_V}
 \end{equation}
which stems from the formula $\Sigma^2(L) = L - \int_0^{L}\mathrm{d}E(L-E)(1-R_2(E))$ and the fact that the 
two-level correlation function $R_2(E)$ for wSRPM is a linear combination of two-level functions of SRPMs 
$\mathcal P_{h_i}^{\beta_i}$.

\section{Level spacing distribution and  number variance 
across the MBL crossover}

The wSRPM model, defined by \eqref{eq: wSRPM} depends on a large number of parameters -- one needs to 
specify JPDFs of the SRPMs $\mathcal P_{\beta_i}^{h_i}$ which contribute to the full JPDF of 
the generalized model $\mathcal P_{wSRPM}$ and find appropriate weight coefficients $c_i$. 
To complete this task we utilize the $P(r_S)$ distributions which encode the inter-sample randomness
across the MBL transition. Distributions of $r_S$ for individual SRPMs $P^{h_i}_{\beta_i}(r_S)$ are  
Gaussian centered around $\overline r^{h_i}_{\beta_i}$ which depends on $h_i$
and $\beta_i$ parameters. The corresponding distribution for wSRPM reads $P_{wSRPM}(r_S) = \sum_i
c_i P^{h_i}_{\beta_i}(r_S)$ and the {set of parameters} $\{ h_i$, $\beta_i, c_i\}$ {is fixed by the requirement that}
$P_{wSRPM}(r_S)$ reproduces the $P(r_S)$ distribution for a {given physical model} (in this case XXZ spin chain \eqref{eq: XXZ})
most faithfully. 
  To fulfil the requirement in a robust way, we select a sequence of coefficients 
$\{(\beta_i, h_i)\}$ with $\overline r^{h_i}_{\beta_i}$ covering the interval $r_S\in[0.386,0.531]$
possibly uniformly, i.e. we choose 
\begin{equation}
 (\beta_i, h_i)=
\begin{cases}
( \frac{i}{100}, 1), \quad  \quad \quad i\in[0,10], \\ 
( \frac{i-8}{20}, 1), \quad  \quad \quad i\in[11,30], \\ 
(1, i-30),\quad  \quad  \, i\in[31,30+h_{\max}],
\end{cases}
 \label{c12}
\end{equation}
where $h_{\max}$ specifies the maximal range of interactions in the contributing SRPMs.
The chosen set of coefficients \eqref{c12} determines the family of wSRPM that can be obtained by various choices of
the weight coefficients $\{c_i\}$.
One way of finding the weight coefficients $\{c_i\}$ would be 
to choose a number of points $r_j$ and solve the linear system of equations $P(r_S=r_j) = P_{wSRPM}(r_S=r_j)$ for
the $\{c_i\}$ coefficients
given that the function $P(r_S=r_j)$ as well as $P^{h_i}_{\beta_i}(r_S)$ are known. If the number of the points 
$r_j$ 
is equal to number of coefficients $c_i$, the solution is unique. Unfortunately, this linear
problem is badly conditioned, small changes of positions of $r_j$ modify drastically the solution $\{ \tilde{c}_i \}$.
In particular, it may happen that certain coefficient $\tilde{c}_k$ is large and positive whereas the next one 
$\tilde{c}_{k+1}$ is large negative
which illustrates that further constraints should be imposed on $\{c_i\}$. Namely, the weight coefficients must be positive 
$c_i>0$ and the differences between subsequent $c_i$ should not be too large. This leads us to a fitting procedure which determines 
the weight coefficients $\{ c_i \}$ by minimizing
\begin{equation}
 \chi^2 = \sum_{i=0}^{i_{\max}} \frac{1}{m} (c_{i+1} - c_i)^2 + \int_{r^m_S}^{r^M_S} (P_{wSRPM}(r_S) - P(r_S) )^2 \mathrm{d}r_S.
 \label{c13}
\end{equation}
 The term $\sum_{i=0}^{i_{\max}} \frac{1}{m} (c_{i+1} - c_i)^2 $ assures the 
``continuity'' of $c_i$ coefficients, the condition $c_i>0$ becomes a constraint of the minimization procedure. 
The constant $m$ is taken as $10^{-4}$ although changing it by a factor of $5$ only mildly affects the results. 
Finally, the minimization of \eqref{c13} does not resolve accurately SRPMs with large $h$ as 
the spacing between subsequent $\overline r^{h}_{\beta}$ decreases drastically, so that we take $h_{\max}=5$
in \eqref{c12}.

Determining the constitutive SRPMs by the choosing the set $\{(\beta_i, h_i)\}$ according to \eqref{c12} and specifying 
the method of obtaining the weight coefficients $\{c_i\}$, we defined a  method of finding wSRPM which 
reproduces inter-sample randomness and can be used across the ergodic-MBL crossover.

\subsection{ Level statistics as function of disorder strength }
\label{subsecdiso}

\begin{table}
\begin{tabular}{S || S S | S S } \toprule
    $W$   & {$\overline r$} & {$ \chi $} & {$ \quad  \quad \overline r_{wSRPM}$} & {$  \quad \quad \chi_{wSRPM}$ }\\ \midrule
    1.5  & 0.5306 & 0.0975 & 0.5282 & 0.078 \\
    1.9  & 0.5219 & 0.259 & 0.5189 & 0.238 \\
    2.1  &  0.5092 & 0.358 & 0.5074 & 0.338\\
    2.5  &  0.4720 & 0.523 & 0.4719 & 0.562 \\ \midrule
    2.9  &  0.4390 & 0.638 &  0.4384 & 0.711 \\
    3.5  & 0.4107 & 0.774 & 0.4101 & 0.850\\
    4.5 & 0.3938 & 0.857 & 0.3941 & 0.929\\\bottomrule
\end{tabular}
 \caption{ \label{tab} {
 Values of the mean gap ratio $\overline r$ and spectral compressibility $\chi$ for XXZ spin chain 
 are compared with predictions of wSRPM model: $\overline r_{wSRPM}$ and $\chi_{wSRPM}$. }
 } 
\end{table}
\begin{figure}
\includegraphics[width=1\columnwidth]{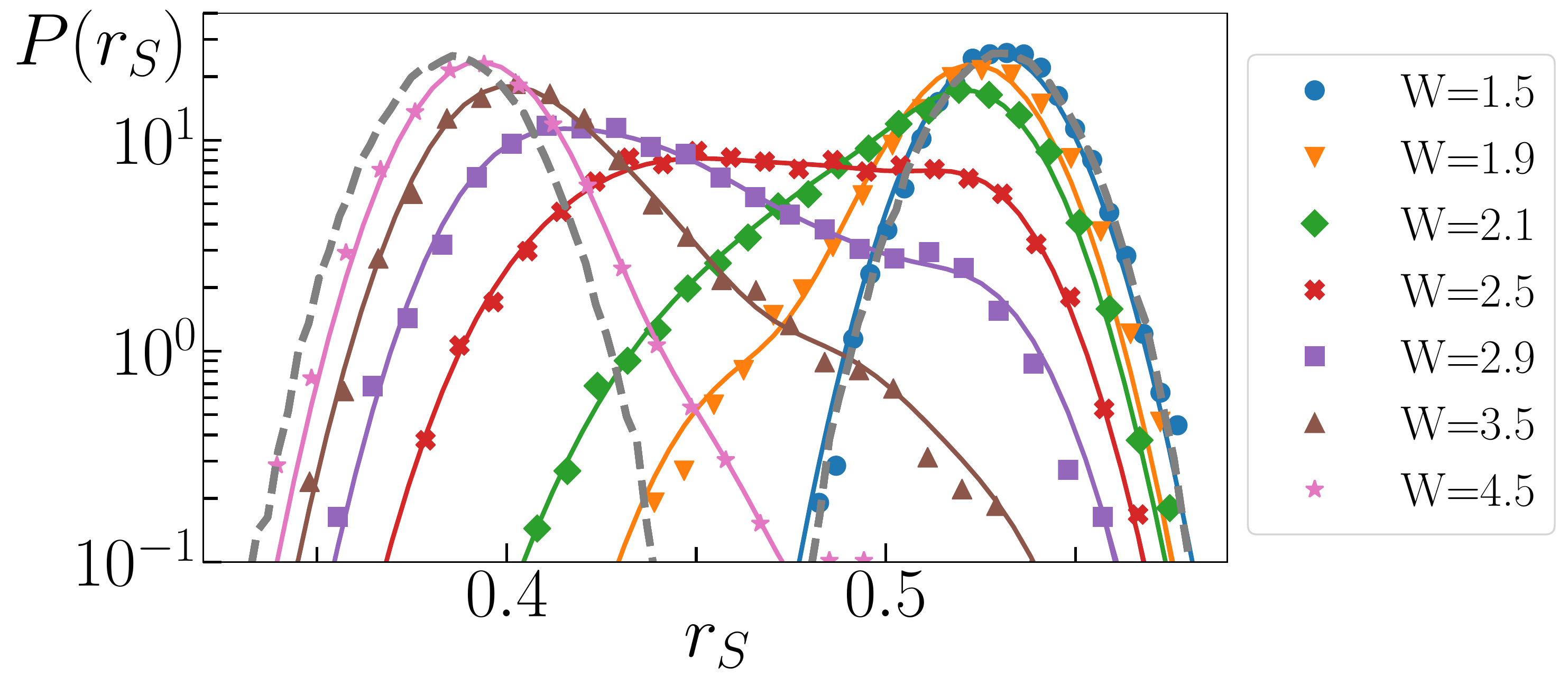}
\caption{ \label{fig: SRPM0} The fit of $P(r_S)$ distributions. Distributions 
$P(r_S)$ of the sample-averaged gap ratio $r_S$ for 
the XXZ spin chain Eq.~\eqref{eq: XXZ} are denoted by markers. 
The corresponding wSRPM fits are denoted with solid lines. {Gray dashed lines correspond
to $P(r_S)$ for GOE and PS center respectively around $\overline{r}_{GOE}=0.531$ and $\overline{r}_{PS}=0.386$.}}
\end{figure}
\begin{figure}
\includegraphics[width=1\columnwidth]{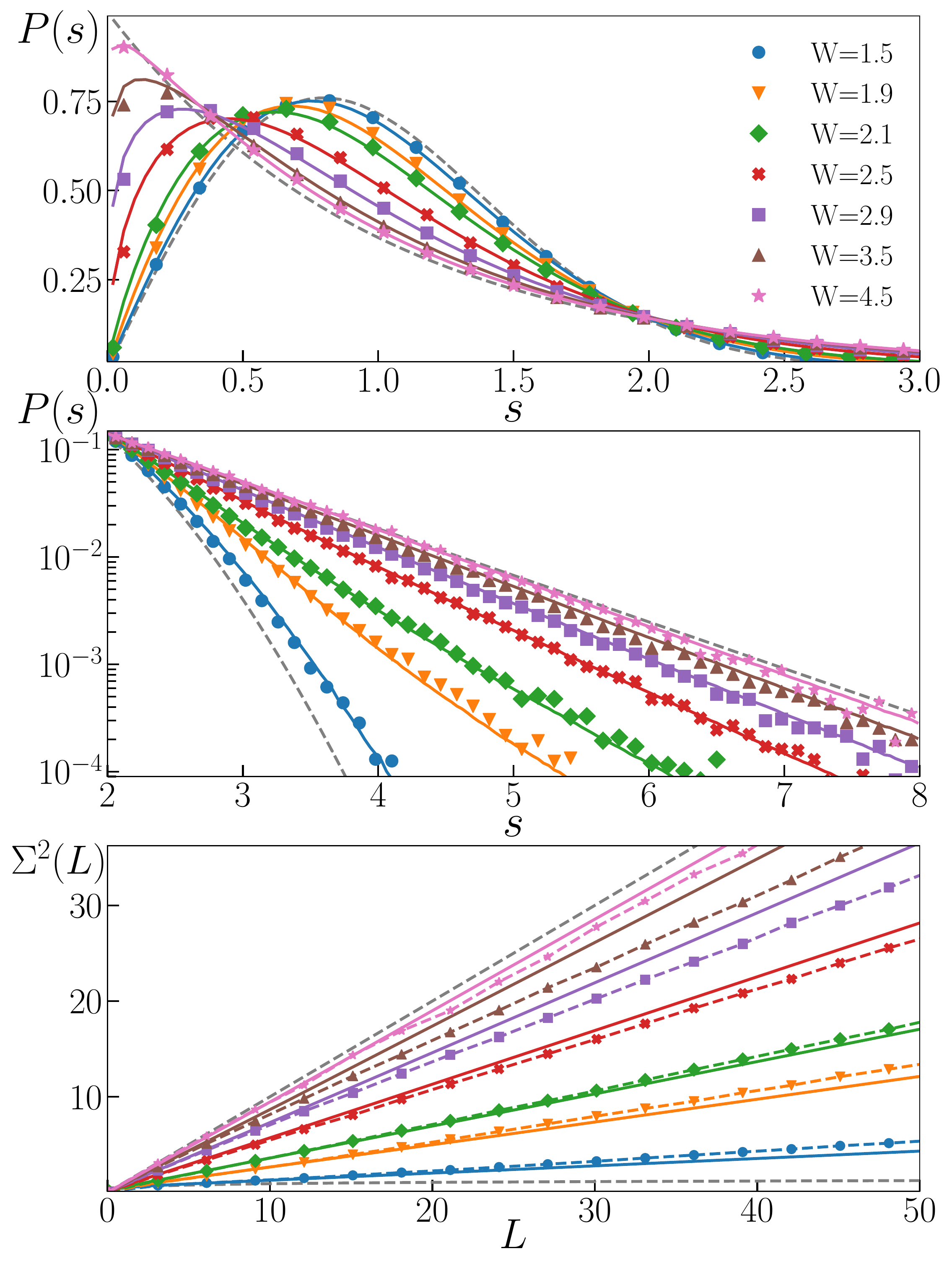}
\caption{ \label{fig: SRPM1} Top panel: level spacing distributions
$P(s)$ {as a function of disorder
strength $W$} in XXZ spin chain \eqref{eq: XXZ} of size $K=16$  are 
denoted by markers,
wSRPM results denoted by solid lines and gray dashed lines denote the level spacing distributions in 
the limiting GOE and PS cases;
Middle panel: same as above, but vertical axis in logarithmic scale to facilitate 
comparison between tails of level spacing distributions;
Bottom panel: dashed lines with markers -- number variance $\Sigma^2(L)$ for XXZ spin chain, solid lines --
results for wSRPM model. {Gray dashed lines correspond
to level spacing distribution or number variance for GOE and PS.} }
\end{figure} 
\begin{figure}
\includegraphics[width=1\columnwidth]{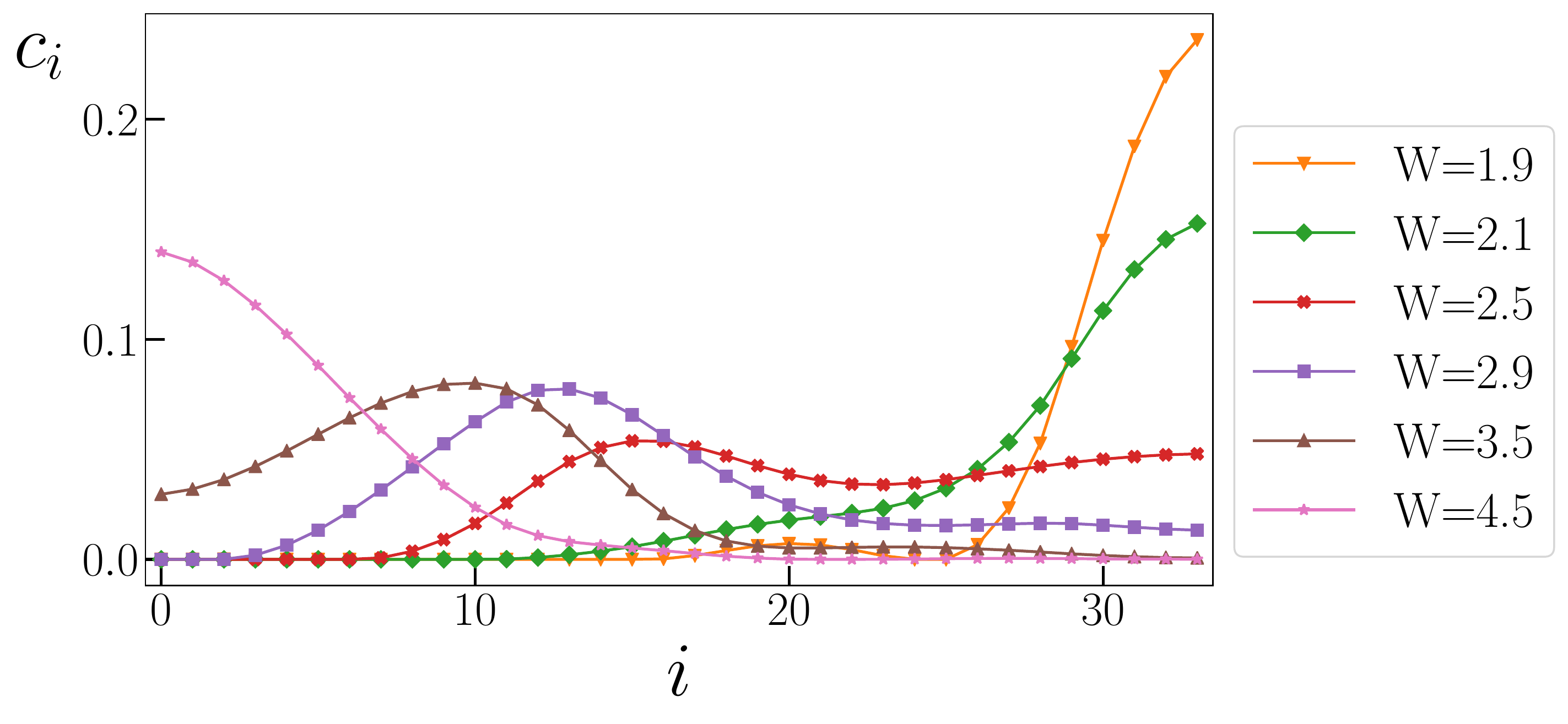}
\caption{ \label{fig: ci1} {Coefficients $c_i$ for the wSRPM for XXZ spin chain across the ergodic -- MBL crossover.
} }
\end{figure}

{The distributions $P_{wSRPM}(r_S)$ found for 
the XXZ spin chain for varying disorder strength $W$ for system of size $K=16$} are presented in Fig.~\ref{fig: SRPM0}. 
Distributions $P(r_S)$ are indeed
recovered in the whole crossover region. {The distribution for $W=1.5$ which is
already very close to the GOE regime is modeled by a single SRPM with $h=12$ and $\beta=1$. }

{Level statistics predicted by the} 
wSRPM model {together with} XXZ spin chain \eqref{eq: XXZ} {data} across the MBL transition
are presented in Fig.~\ref{fig: SRPM1}. We have accumulated data for $n=2000$ disorder realizations 
for each disorder strength $W$ and we have set the mean level spacing to unity (details described in the 
Sec. \ref{long-range-section}). The solid lines which denote the predictions of wSRPMs 
match with a very good accuracy both the bulks and the tails of level spacing distributions for disorder 
strengths $W$ corresponding to the whole regime intermediate between GOE and PS level statistics.
In particular, the tails of $P(s)$ for $W=1.9, 2.1$ are visibly bent upwards -- this is a clear 
manifestation that SRPMs which account for more localized rare events must be included in the wSRPM.
The number variances $\Sigma^2(L)$ predicted by the wSRPM are again reproducing the data for 
XXZ spin chain \eqref{eq: XXZ} with a very good precision. The number variance predicted by fitting of a single SRPM 
presented in Fig.~\ref{fig: cmp} was underestimating the result for {$W=1.9$} and it is the 
contribution from other SRPMs included in the wSRPM which ensures the agreement in the number variance.

Specific values of the weight coefficients are presented in Fig.~\ref{fig: ci1}.
We also compare the mean gap ratio $\overline r$ with the prediction of wSRPM 
$\overline r_{wSRPM} = \sum_i c_i \overline r^{h_i}_{\beta_i}$ in Tab.~\ref{tab} showing the agreement
at the level of $0.5\%$. In addition, we also collate spectral comprehensibilities. Predictions
of wSRPM $\chi_{wSRPM} =  \sum_i c_i /(h_i \beta_i +1)$ agree with spectral comprehensibilities
for XXZ spin chain obtained from quadratic fit to the number variance 
$\Sigma^2(L)$ for XXZ spin chain in the interval $L\in[10,70]$ (see also Section~\ref{long-range-section})
up to $10\%$.

\subsection{ Level statistics as a function of system size $K$}
\begin{table}
\begin{tabular}{S || S S | S S } \toprule
      & {$\overline r$} & {$ \chi $} & {$ \quad  \quad \overline r_{wSRPM}$} & {$  \quad \quad \chi_{wSRPM}$ }\\ \midrule
      
      {$W=2.1$} \\ 
    {$K=14$}  & 0.4956& 0.392 & 0.4946 & 0.432 \\
    {$K=16$}  &  0.5092 & 0.354 & 0.5074 & 0.338\\
    {$K=18$}  &  0.5221 & 0.259 & 0.5193& 0.231 \\ \midrule
    {$W=4.0$} \\ 
    {$K=14$}   &  0.4021 & 0.743 &  0.4024 & 0.888 \\
    {$K=16$}   & 0.3996 & 0.805 & 0.3992 & 0.905\\
    {$K=18$}  & 0.3989 & 0.849 & 0.3981 &  0.910\\\bottomrule
\end{tabular}
 \caption{ \label{tab2} {
 Values of the mean gap ratio $\overline r$ and spectral compressibility $\chi$ for XXZ spin chain 
 are compared with predictions of wSRPM model: $\overline r_{wSRPM}$ and $\chi_{wSRPM}$. }
 } 
\end{table}

\begin{figure}
\includegraphics[width=1\columnwidth]{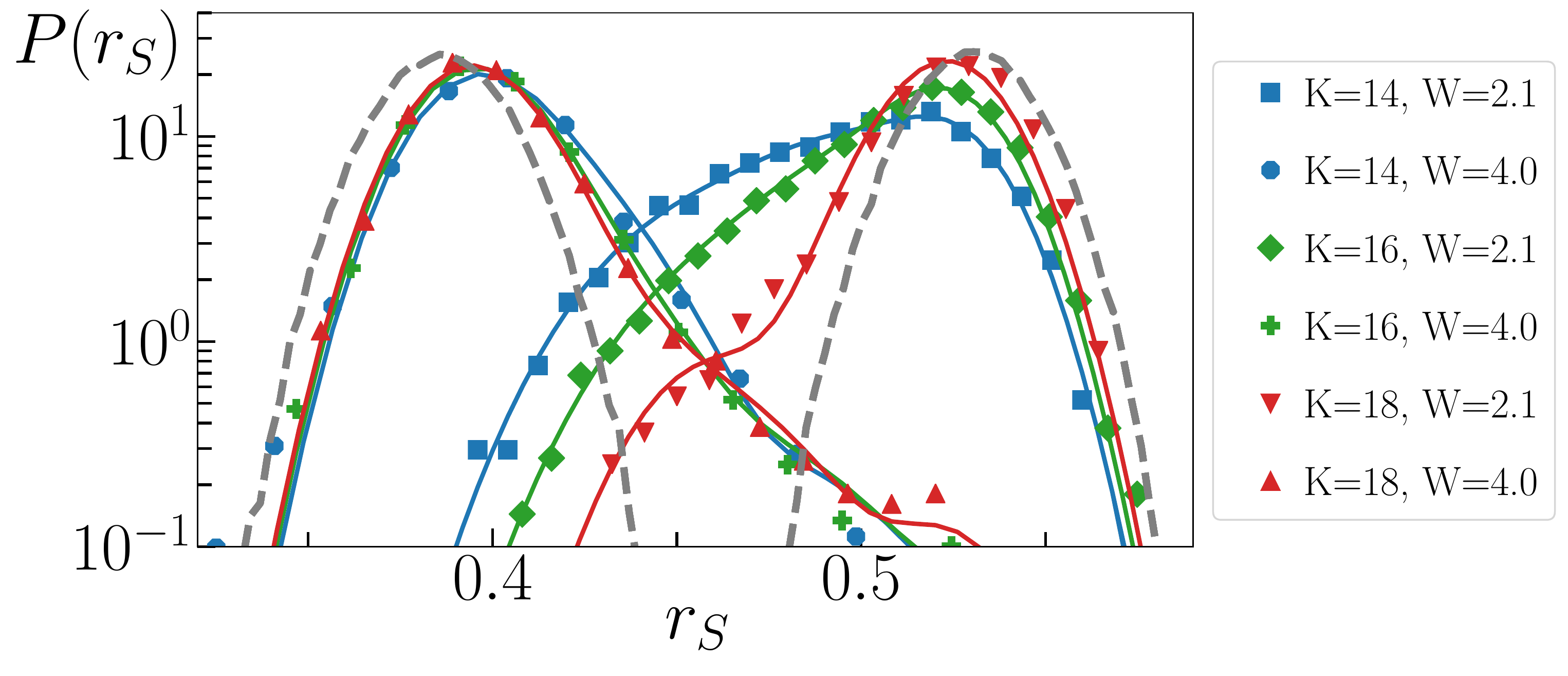}
\caption{ \label{fig: SRPM2a} The fit of $P(r_S)$ distributions. Distributions 
$P(r_S)$ of the sample-averaged gap ratio $r_S$ for 
the XXZ spin chain Eq.~\eqref{eq: XXZ} are denoted by markers. 
The corresponding wSRPM fits are denoted with solid lines. {Gray dashed lines correspond
to $P(r_S)$ for GOE and PS center respectively around $\overline{r}_{GOE}=0.531$ and $\overline{r}_{PS}=0.386$.}}
\end{figure}
\begin{figure}
\includegraphics[width=1\columnwidth]{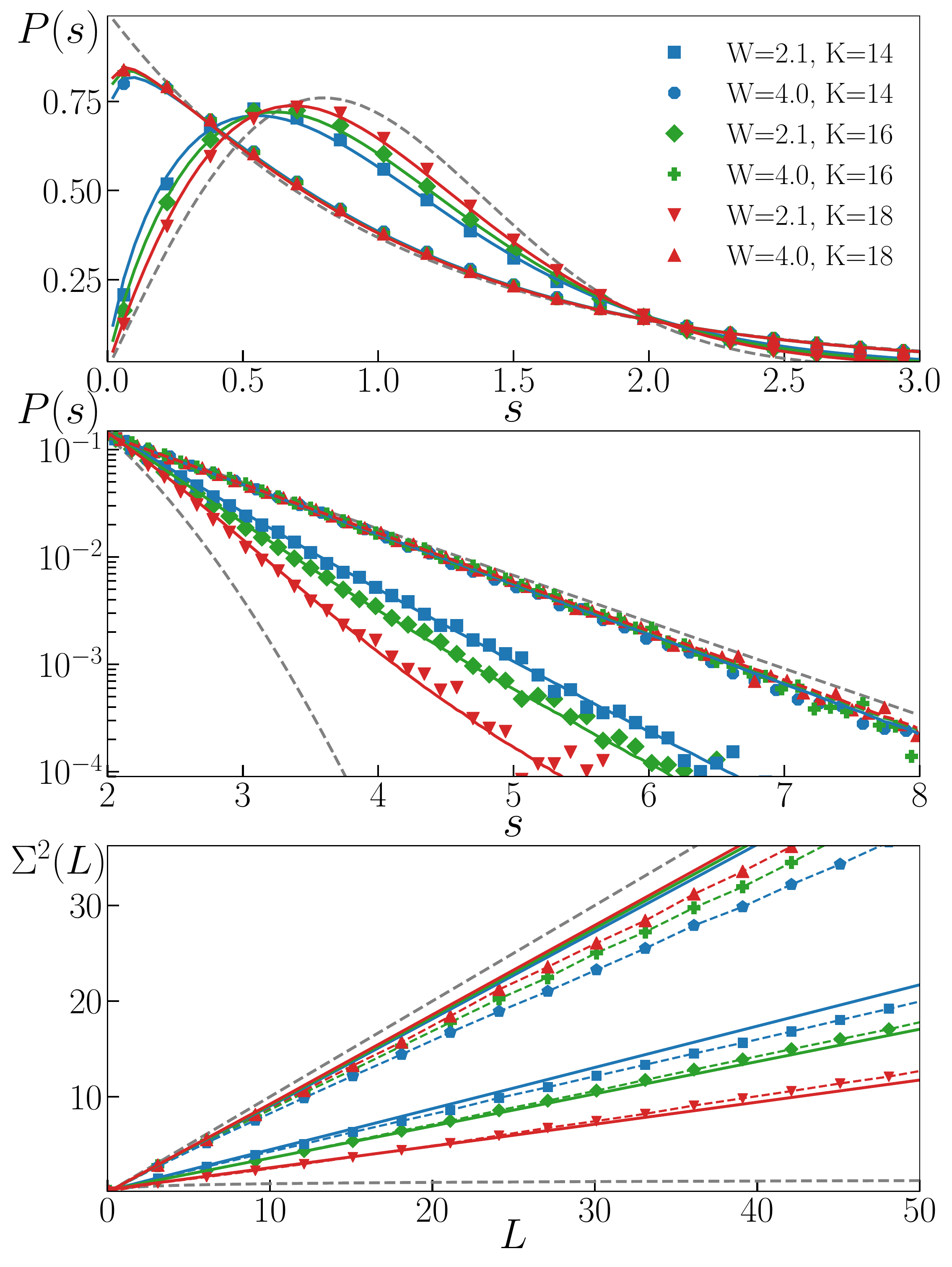}
\caption{ \label{fig: SRPM3a} Top panel: level spacing distributions
$P(s)$  in XXZ spin chain \eqref{eq: XXZ} {for varying system size $K$}   are 
denoted by markers,
wSRPM results denoted by solid lines and gray dashed lines denote the level spacing distributions in 
the limiting GOE and PS cases;
Middle panel: same as above, but vertical axis in logarithmic scale to facilitate 
comparison between tails of level spacing distributions;
Bottom panel: dashed lines with markers -- number variance $\Sigma^2(L)$ for XXZ spin chain, solid lines --
results for wSRPM model. Gray dashed lines correspond
to level spacing distribution or number variance for GOE and PS. }
\end{figure} 
\begin{figure}
\includegraphics[width=1\columnwidth]{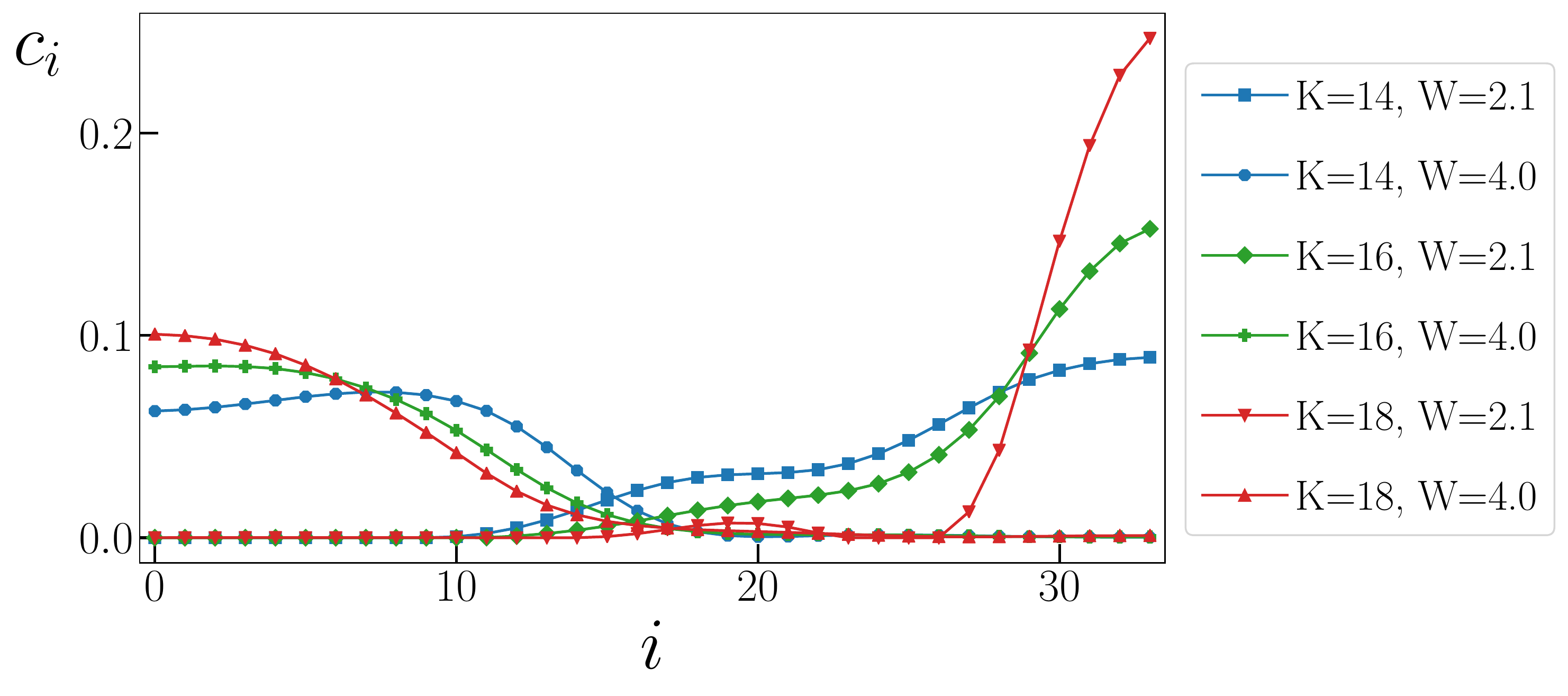}
\caption{ \label{fig: ci2a} {Coefficients $c_i$ for the wSRPM for XXZ spin chain {for varying system size on the two 
sides of } the ergodic -- MBL crossover.
} }
\end{figure}

{We have thus demonstrated that by constructing appropriate wSRPMs one can model flow of level statistics
across the ergodic-MBL crossover for XXZ spin chain at given system size. In the following subsection 
we demonstrate how the level statistics evolve with system size $K$ for fixed disorder strength $W$.}

{A quick glance at Tab.~\ref{tab2} reveals the average gap ratio $\overline r$ tends towards $\overline
r_{GOE}$ and $\overline r_{PS}$  values with increasing system size $K$
at disorder strengths close either to the ergodic regime ($W=2.1$) or to the MBL phase ($W=4.0$).
The system size dependence is much stronger in the former case than in the latter as it is also visible in 
Fig.~\ref{fig: SRPM2a}. The evolution of level spacing distribution $P(s)$ and number variance 
$\Sigma^2(L)$ with increasing 
system size is presented in Fig.~\ref{fig: SRPM3a}. The level spacing and the number variance change with 
increasing size as the gap ratio suggest flowing either to GOE or to PS limit. The wSRPM 
determined by the requirement of reproducing the the $P(r_S)$ distributions shown in Fig.~\ref{fig: SRPM2a}
captures accurately level spacing distribution as well as the number variance for the considered system sizes. 
The evolution of parameters
of the model with increasing system size is presented in Fig.~\ref{fig: ci2a} showing clearly that the weight of
SRPMs closer to the GOE (those with larger $i$) increases with the growing system size. Analogous (albeit weaker) dependence
is observed also close to the MBL regime ($W=4.0$).
}

\subsection{ Flow of level statistics in the ergodic-MBL crossover}

We have shown that wSRPM accurately describes level statistics of disordered XXZ spin chain across the 
whole ergodic to MBL crossover. The agreement is remarkably good even for long-range spectral correlations
as shown by the number variance $\Sigma^2(L)$. This leads us to conclusion that the effective interactions
between eigenvalues in the MBL crossover are accurately grasped by level correlations of wSRPM.

The picture of  the flow of level statistics from GOE to PS in the MBL transition which emerges 
is the following. In the ergodic phase the range of interactions between eigenvalues tends to infinity, $h=\max \{h_i\} 
\rightarrow \infty$, and the level statistics
reduces to GOE case. As the disorder strength increases, the range of interactions 
between eigenvalues $h$ declines to a finite value, level spacing distribution acquires an exponential tail
and a finite spectral compressibility $\chi$ appears as the number variance grows linearly $\Sigma^2(L) \propto \chi L$. 
Upon further increase of the disorder strength, the range of interactions
$h$ decreases further.
A larger contribution of level statistics with short-range interactions 
appears as it is visible in tails of
level spacing distributions
and in the enhancement of spectral compressibility $\chi$. As the MBL phase is approached the interactions
become local $h=1$ and parameter $\beta=\max\{\beta_i\}$ starts to flow from $\beta=1$ to $\beta=0$ in the MBL 
phase similarly as in the second stage of the flow described in \cite{Serbyn16}. This final stage of the 
flow is also accompanied by rare inclusions of systems which have nearly ergodic properties {as} it is 
visible in the
$P(r_S)$ distribution in Fig.~\ref{fig: SRPM0}.
The presence of this contribution also slightly diminishes the number variance.

Level statistics on the ergodic side of crossover flow towards GOE limit for growing system size. Similarly, 
for disorder strength $W \gtrsim 4.0$ system approaches the PS limit as its size increases. 
The transition becomes sharper and the
range of disorder strengths $W$ for which a wSRPM with more than a single non-zero coefficient
shrinks with increasing system size. We speculate 
that in the $K \rightarrow \infty$ limit only at the critical disorder strength $W_C$ the level statistics is
neither GOE nor PS. Candidate for such critical level statistics for the MBL transition is discussed in 
the subsection \ref{sec: critical}.

In conclusion, the wSRPM allows to model the level statistics in the XXZ spin chain in the whole MBL crossover. 
The level statistics are reproduced with a nearly perfect agreement on the level of ten level spacings.
Slight discrepancies associated with long-range spectral correlations are discussed 
{in Sec.~\ref{long-range-section}}. 
Now, we proceed to discussion of the critical level statistics in subsection \ref{sec: critical} to further 
demonstrate that wSRPM describes also
statistics observed for other systems that reveal MBL transition in Sec.~\ref{boson-section}.

\subsection{Critical level statistics in MBL transition}
\label{sec: critical}

 \begin{figure}
\includegraphics[width=1\columnwidth]{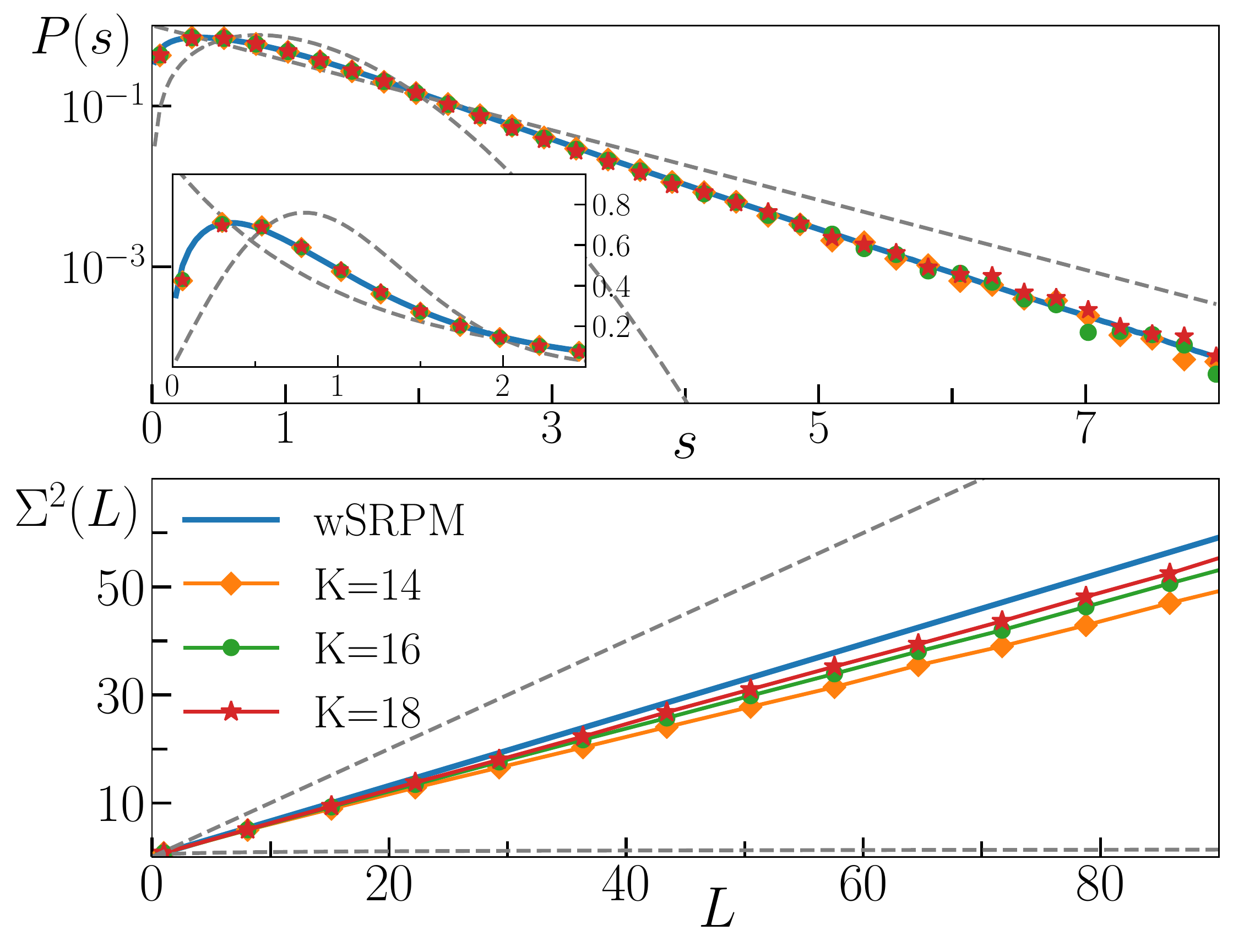}
\caption{ \label{fig: crit} Critical level statistics for XXZ spin chain with random uniform disorder.
{Dashed lines correspond to the GOE and Poisson cases.}
}
\end{figure} 
\begin{figure}
\includegraphics[width=1\columnwidth]{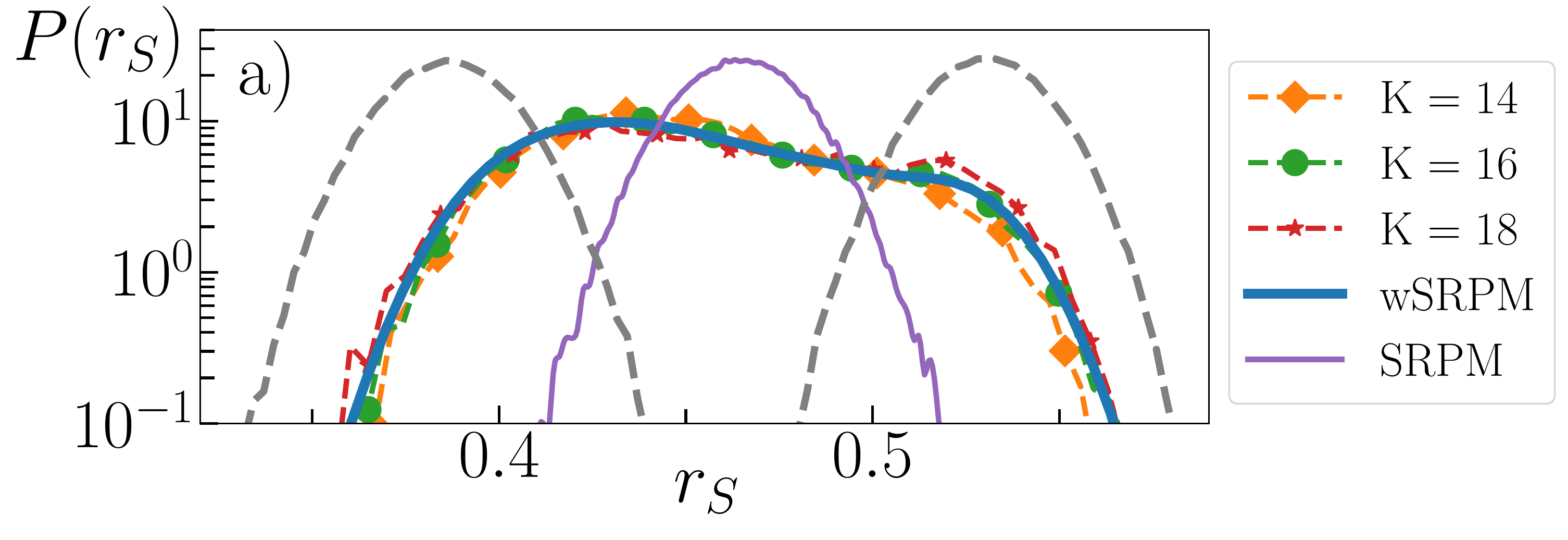}
\includegraphics[width=1\columnwidth]{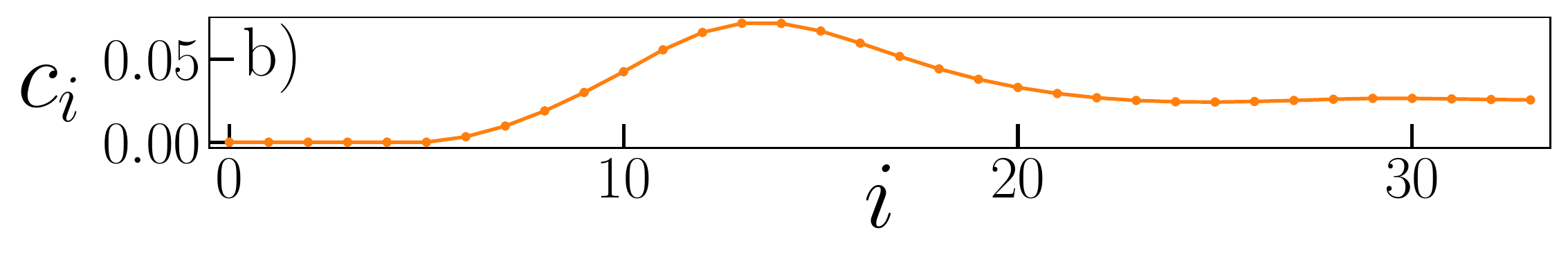}
\caption{ \label{fig: critPRS}  {a)} The $P(r_S)$ distribution
for the critical statistics in XXZ spin chain
along with wSRPM and $P(r_S)$ for SRPM. b) The $c_i$ coefficients of wSRPM  shown in  Fig~\ref{fig: crit}.
}
\end{figure} 
 \begin{table}
\begin{tabular}{ S S || S S }  \toprule
  $K$ &  $W$  & {$\overline r$} & {$ \chi $} \\ \midrule
   14 & 2.62  &  0.4528(4) & 0.545(9) \\
   16 & 2.7   &  0.4537(5)& 0.587(5) \\
   18 & 2.8   &  0.4569(7) & 0.605(4) 
 \\\bottomrule
   & & {$\overline r_{wSRPM}$} & {$ \chi_{wSRPM}$ } \\ \cmidrule{3-4}
   &    &  0.4530& 0.639 \\ \cmidrule[0.8pt]{3-4}
\end{tabular}
 \caption{ \label{tab2}  The average gap ratio $\overline r$ and spectral compressibility $\chi$
 for the XXZ spin chain at disorder strength which corresponds to $W_C$ at in the thermodynamic limit 
 $K\rightarrow \infty$.
 For comparison, the predictions of wSRPM $r_{wSRPM}$ and $ \chi_{wSRPM}$ are displayed.
 } 
\end{table}

 We assume that the critical level statistics in MBL transition 
can be extracted from data for a system of size $K$ for disorder strength $W_K$ that maximizes the inter-sample
variance $V_S$, e.g. $W_K=2.7$ for $K=16$. The finite size analysis assures that in the thermodynamic 
limit $K\rightarrow \infty$: $W_K \rightarrow W_C =3.5(1)$ (as discussed in Section~\ref{sec: ranal}).

As the system size increases the statistics on the ergodic (MBL) side of crossover tend towards GOE (Poisson) limit, the 
width of the crossover diminishes. 
 The critical level statistics which we conjecture to be relevant exactly at the MBL transition 
 in large system size limit is presented in Fig.~\ref{fig: crit}. 
 {The obtained wSRPM contains SRPMs with long-range interactions $h>1$ (non-zero weights $c_i$ with $i>30$) together with 
 dominating contribution of models with local interactions and $\beta< 1$. Large number of contributing SRPMs allows to 
 accurately reproduce the $P(r_S)$ distribution (Fig.~\ref{fig: critPRS}).
 Moreover, it is vital
 to faithfully reproduce the number variance. 
 The values of spectral compressibility $\chi$ defined by the linear large $L$ behavior of the 
 number variance $\Sigma^2(L) \propto \chi L$ together with the average gap ratios $\overline r$
 are shown in Tab.~\ref{tab2}. This quantities are in good agreement with the 
 predictions of the wSRPM $\overline r_{wSRPM}$ and $\chi_{wSRPM}$. 
  The data suggest that the remaining small deviation in the spectral compressibility
  $\chi$
  is probably a finite size effect.

Level spacing distribution $P_A(s)$ in the Anderson transition \cite{Shklovskii93}
combines level repulsion at small $s$ characteristic for GOE and an exponential tail of Poisson level statistics, the critical statistics
shown in Fig.~\ref{fig: crit} also possess the two features. However, the large inter-sample randomness
encoded in broad $P(r_S)$ distribution is a crucial property of the critical level statistics in MBL transition, whereas 
it does not play a role in the Anderson transition in which the $P_A(r_S)$ distribution has a 
Gaussian shape the same width as in the 
GOE and Poisson limits.

\section{Universality}
\label{boson-section}

\begin{figure*}
\includegraphics[width=0.5\textwidth]{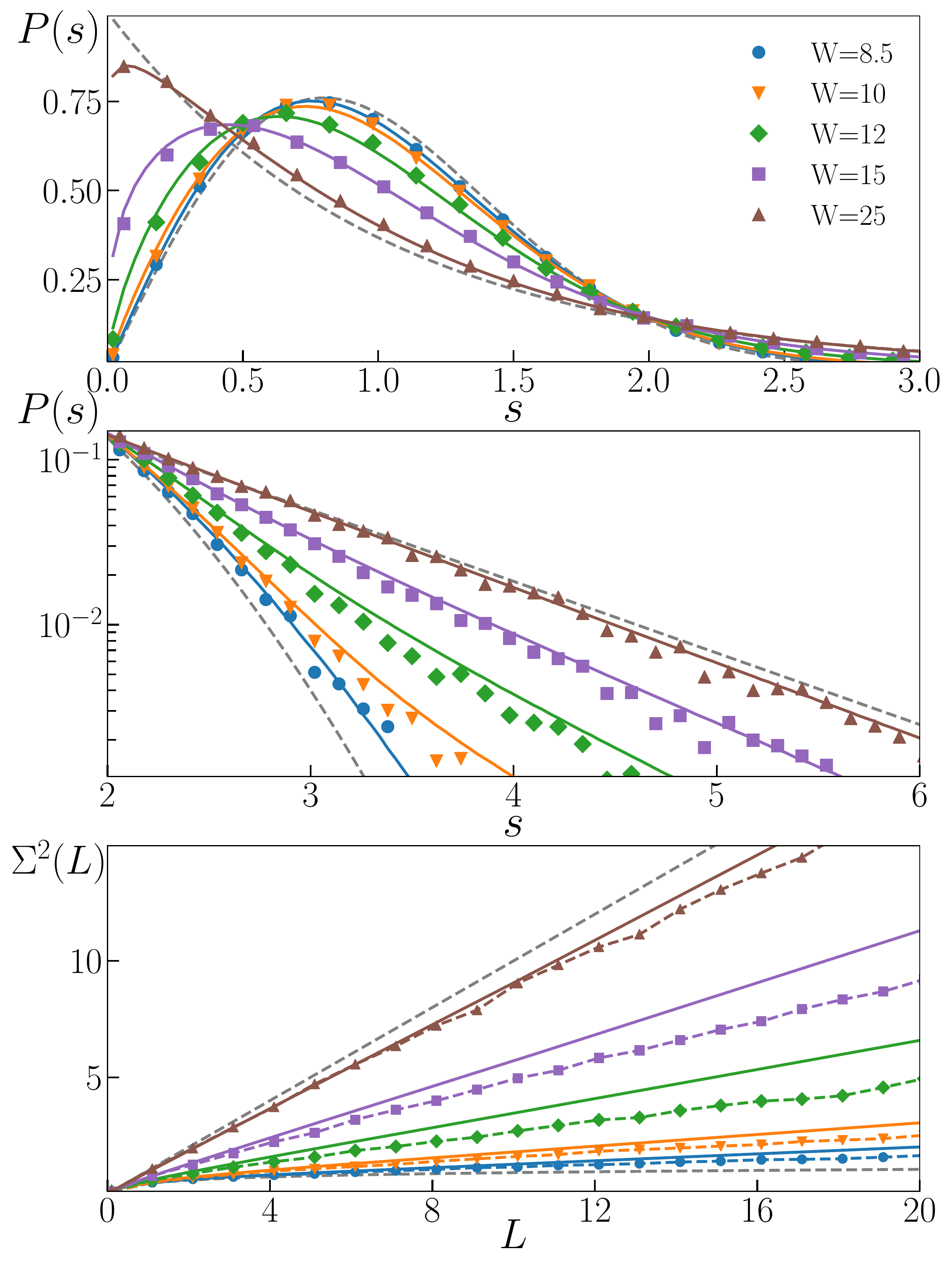}\includegraphics[width=0.5\textwidth]{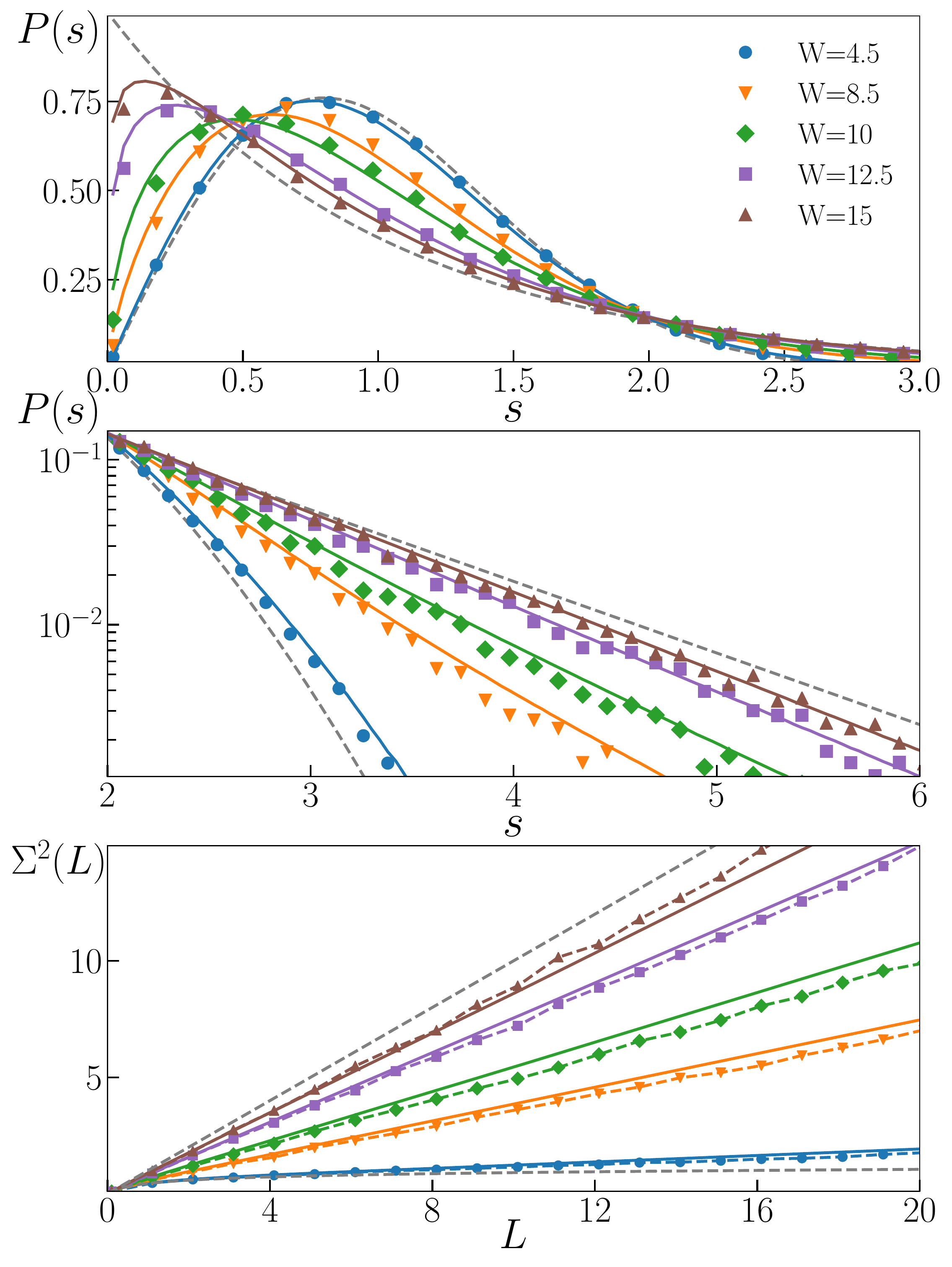}
\caption{ \label{fig: BH} Left: Level spacing distribution $P(s)$ (in lin-lin scale 
and in lin-log scale) together with number variance $\Sigma^2(L)$ during MBL transition in disordered 
Bose-Hubbard model \eqref{eq: ham_BH}. Results for $N=12$ bosons on $K=8$ lattice sites, 
interaction amplitude $U=1$ are denoted by markers, solid lines show wSRPM model fits. 
Right: Level statistics for the Fermi-Hubbard model $H_F$. Results for $N_{\uparrow} = 3 = N_{\downarrow}$
fermions on $K = 12$ lattice sites with interaction strength $U=2$ and $\mu_B=h_B=0.1$, $J'=0.5$
are denoted by markers, solid lines correspond to wSRPM model predictions.}
\end{figure*}

The wSRPM model has so far been used to describe level statistics in the standard model of MBL 
-- the XXZ spin chain \eqref{eq: XXZ}. It has already been noted in \cite{Santos10}
that there are differences in level statistics across the MBL transition in systems 
of hard-core bosons and fermions. In this section we demonstrate that 
wSRPM can {faithfully} reproduce level statistics in 
{ergodic to MBL crossover}
in a  disordered Bose-Hubbard model \cite{Sierant18} as well as in disordered
Fermi-Hubbard model \cite{Mondaini15}.

\begin{figure}
\includegraphics[width=1\columnwidth]{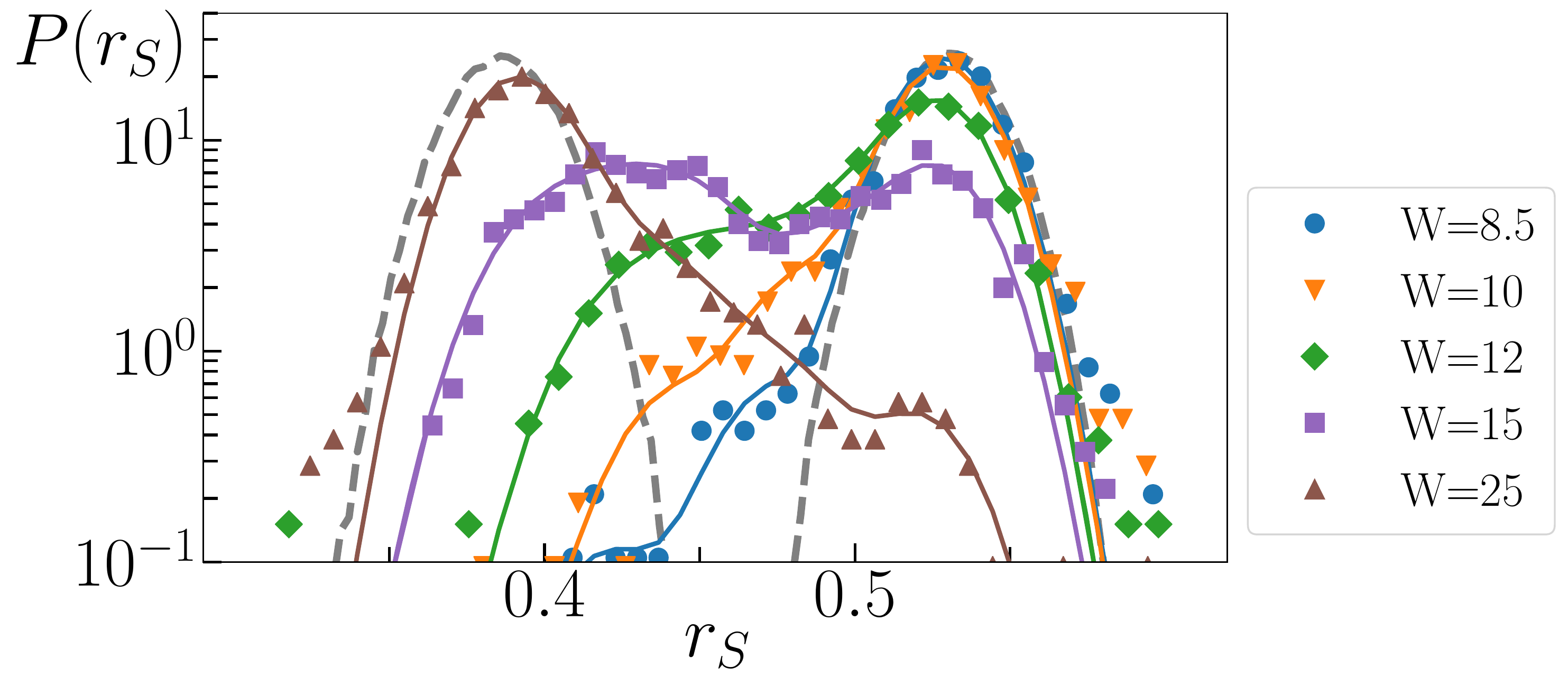}
\caption{ \label{fig: BBPRS} {The fit of $P(r_S)$ distributions. Data for bosonic system 
\eqref{eq: ham_BH} are denoted with markers, solid lines show wSRPM fits.} }
\end{figure}

The system of disordered bosons is described by the Bose-Hubbard Hamiltonian
\begin{equation}
  H_{B} = -J \sum_{\langle i,j \rangle} \hat{a}^{\dag}_i\hat{a}_j +
 \frac{ U }{2} \sum_i \hat{n}_i (\hat{n}_i - 1) +
  \sum_i \mu_i \hat{n}_i,
 \label{eq: ham_BH}
\end{equation}
where  $a^{\dag}_i, a_i$ are bosonic creation and annihilation operators respectively, 
the tunneling amplitude $J=1$ sets the energy scale, $U$ is interaction strength and the chemical 
potential $\mu_i$ is distributed uniformly in an interval $[-W;W]$. This model have
been shown to be MBL \cite{Sierant18} above a critical disorder 
strength $W_B$ which depends on the interaction strength $U$. 
The Hamiltonian for disordered fermions reads
\begin{equation}
  H_{F0} = -J \sum_{i,\sigma=\uparrow, \downarrow}( \hat{c}^{\dag}_{i\sigma}\hat{c}_{i+1\sigma} +h.c)+
  U\sum_i n_{i\uparrow}n_{i\downarrow} +
  \sum_i \mu_i \hat{n}_{i}
 \label{eq: ham_FH}
\end{equation}
where $c^{\dag}_i c_i$ are fermionic creation and annihilation operators respectively; $J=1$ and $U$
are tunneling and interaction amplitudes and $\mu_i \in [-W;W]$ is uncorrelated disorder. 
To avoid integrability in the absence of disorder it is sufficient \cite{Mondaini15} to add  
the next-to-nearest neighbor tunneling terms 
\begin{equation}
  H_1 = -J' \sum_{i,\sigma}( \hat{c}^{\dag}_{i\sigma}\hat{c}_{i+2\sigma} +h.c)
\end{equation}
and an additional symmetry breaking term 
\begin{equation}
  H_{SB} = h_B(n_{i\uparrow}-n_{i\downarrow}) + \mu_B (n_{L\uparrow}+n_{L\downarrow}).
\end{equation}
Transition between GOE and PS statistics for the system with 
the full Hamiltonian 
\begin{equation}
H_F=H_{F0} + H_1+H_{BS}
\label{eq: HF}
\end{equation}
has been observed in \cite{Mondaini15}.

Level statistics {as a function of 
disorder strength in}  the bosonic \eqref{eq: ham_BH} and fermionic \eqref{eq: HF}
models together with wSRPM fits are presented in Fig.~\ref{fig: BH}.
Similarly as in the case of XXZ spin chain, the level spacing distributions $P(s)$ 
are characterized by exponential tails (which are also bending upwards for large $s$), the number 
variance $\Sigma^2(L)$ is growing linearly at large $L$ similarly as in the case of disordered XXZ chain. 
 It seems that these are universal features 
of level statistics in { ergodic to MBL crossover  in models with short-range interactions. Such systems
host an extensive number of LIOMs. Presumably, the observed common features of level statistics 
across the MBL crossover are associated with the way in which the LIOMs get delocalized as the disorder strength decreases.}

{Predictions of wSRPM reproducing appropriate $P(r_S)$ distributions for both systems are
denoted by solid lines in Fig.~\ref{fig: BH}. In the case
of the fermionic system, the method of determining the wSRPM model was exactly the same as for the XXZ spin chain, 
whereas for bosons we have also included SRPM with $\beta=1$ and $h=25$ (it has $\overline r^{h=25}_{\beta=1} = 0.5302$  
which can be compared with $r^{h=5}_{\beta=1} = 0.5262$). For the smallest disorder strengths $W=8.5$ ($W=4.5$) 
for bosons (fermions), a single SRPM with $h=12$ ($h=13$)  was fitted. The bulks of level spacing distributions as well 
as the tails are reproduced reasonably accurately by the wSRPM across the whole MBL crossover. The wSRPM prediction
for the number variance $\Sigma^2(L)$ is compatible with data for disordered Fermi-Hubbard model. However,
the number variance is significantly $20-30\%$ overestimated for the disordered Bose-Hubbard model at $W=10$ and $W=12$.
Inspecting closely the $P(r_S)$ distributions in Fig.~\ref{fig: BBPRS} we can clearly see abundance of
disordered realizations with $r_S\gtrsim 0.57$ for $W=10, 12$ such that $P(r_S)$ is above the GOE distribution.
Precisely this abundance lead us to consider also the  $\beta=1$ and $h=25$ SRPM in the fits for bosonic system.
While it diminishes the deviation of the number variance, it is clearly insufficient to yield correct spectral 
compressibility.
}
This {demonstrates} 
model specific {long-range correlations between eigenvalues {that} cannot be grasped straightforwardly 
by wSRPM}. 

{Apart from the model specific details,}
the exponential tails of level spacing distributions and the finite spectral compressibility
that appear already deeply in the metallic phase 
were observed for the XXZ spin chain as well as in the bosonic and fermionic systems. 
 The wSRPM model is able to grasp all of those features
which provides an argument in favor of its generality.
Moreover, distinct numbers of rare events occur in various systems during MBL transition
which reveals itself in dissimilar correspondences between the bulk of level spacing distribution
and its tail as well as the number variance.
In general, different systems are characterized by different inter-sample randomness
during the MBL transition -- compare for instance the shape of $P(r_S)$ distributions displayed 
in Fig.~\ref{fig: BBPRS} with data for the XXZ spin chain in Fig.~\ref{fig: SRPM0}. 
This demonstrates that 
{an accurate model of level statistics must be flexible enough to reproduce 
various types {of} inter-sample randomness across the MBL crossover -- 
this necessitates an introduction of a weighted model like wSRPM.}

\section{Long-range spectral correlations}
\label{long-range-section}

The number variance $\Sigma^2(L)$ at large $L$ reflects correlations between energy 
levels which lie far apart in the spectrum of a system. Such long-range correlations
between eigenvalues are strong in the GOE ensemble, resulting in the so called spectral rigidity 
which is apparent in the asymptotic behavior of the number variance $\Sigma^2_{GOE}(L) \rightarrow \log(L)$ at $L \gg 1$.
The spectral rigidity of GOE is associated with the fact that the logarithmic interactions
act between all pairs of eigenvalues in 
the JPDF for GOE \eqref{eq: JPDF1a}. And it is the spectral rigidity of $\beta$-Gaussian model which 
causes the large discrepancy between its prediction and the number variance
for XXZ spin chain in Fig.~\ref{fig: cmp}.
On the other hand, the SRPMs describe interactions only among a finite number $h$ of neighboring 
eigenvalues
which results in the spectral compressibility of those models 
$\Sigma^2_{GOE}(L) \rightarrow \chi L$ at $L \gg 1$,
with $0<\chi<1$. The resulting spectral compressibility of the wSRPM model
allows to grasp the linear behavior of 
number variance in the MBL {crossover}. The similar behavior can be also
obtained with wPLRBM as presented in the preceding section.

To be able to compare
statistical properties of eigenvalues from different parts of spectra of various systems, one
has to perform 
the unfolding of energy levels \cite{Haake} -- the procedure of setting mean level spacing to unity.
Unfortunately, the number variance $\Sigma^2(L)$ is very sensitive to details of the unfolding 
\cite{Gomez02}
which has already been a source of discrepancies in descriptions of level statistics in the MBL 
transition 
\cite{Serbyn16,Bertrand16}. 
Consider a set of eigenvalues $\{E_i\}$ ordered in an ascending manner. During the unfolding, a 
level staircase 
function $\sigma(E)= \sum_i \Theta(E- E_i)$ is separated into smooth and fluctuating parts $\sigma(E) = 
\overline \sigma(E) + \delta \sigma (E)$ and the eigenvalues are mapped via 
\begin{equation}
 E_i \rightarrow \epsilon_i
= \overline \sigma(E_i).
\label{eq: unf1}
\end{equation}
The difficulty of unfolding lies in an ambiguity of  the definition of the smooth
part $\overline \sigma(E)$ of the staircase function.  The most common way is to fit 
the staircase function $\sigma(E)$ for each disorder realization with a polynomial of a small degree
{which determines the smooth part} $\overline \sigma(E)$. 

In our case, a set of $n=400$ consecutive eigenvalues is gathered and 
the resulting level staircase is fitted with a straight line which defines the smooth part $\overline \sigma(E)$
used in the unfolding of energy levels. For each disorder realization $7$ non-overlapping sets of $n=400$ eigenvalues 
from the middle of spectrum are taken -- effectively employing $\approx20\%$ of the spectrum to the analysis 
as the matrix
size for $K=16$ is equal to $12870$. 
The finite size $n$ of the set of eigenvalues introduces a correction $-a_2 L^2/n$ to the number variance \cite{Ndawana03}.
Carrying out the unfolding with $n=50,100,200,400,800$ we verify that it is indeed the case. We perform 
a quadratic fit to $\Sigma^2(L)$ in the interval $L \in [10,70]$ and obtain the coefficient  $a_2$ which 
is {weakly dependent on the chosen $L$ interval}. Therefore, in order to eliminate the quadratic correction and thus 
to get rid of the finite $n$ effects we subtract the $-a_2 L^2/n$
term from the number variance data. 
Let us note that unfolding with finite number $n$ of energy levels can have two consequences.
 For eigenvalues which are strongly correlated at large distances (e.g. GOE), 
 it destroys level correlations at approximately  $ n$ 
 level spacings meaning  that at this ranges the eigenvalues become uncorrelated. Hence, the number variance 
 becomes overestimated at $L \approx n$. The converse is true for uncorrelated energy levels -- unfolding based on $n$ 
 energy levels introduces correlations between them at a certain scale -- and the number
 variance is underestimated. We have checked that our unfolding procedure (together with the $-a_2 L^2/n$ term
 subtraction) allows us to get correct number variances in the two limiting cases of GOE and PS statistics 
 up to $L \approx 100$.

\begin{figure}
\includegraphics[width=1\columnwidth]{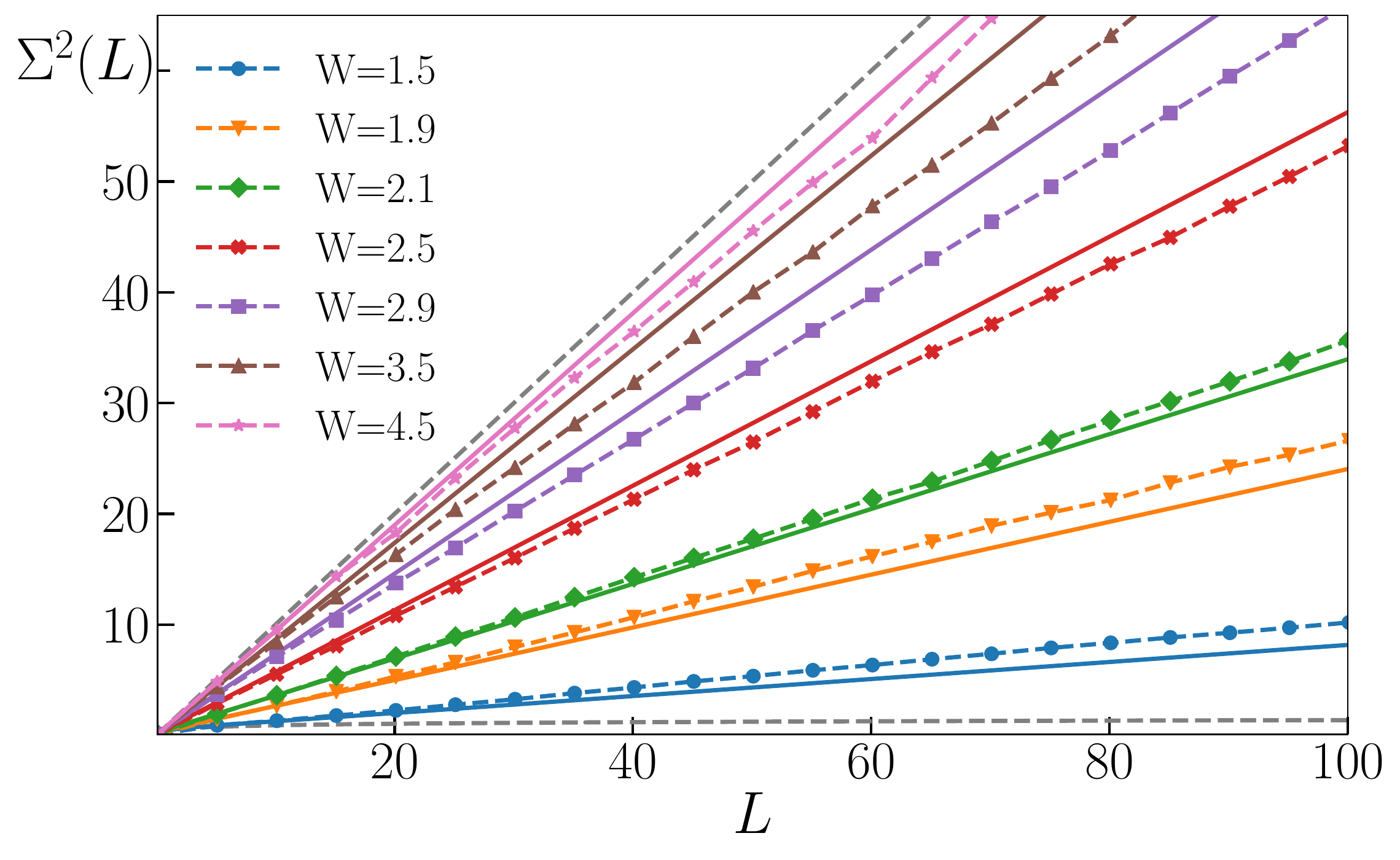}
\caption{ \label{fig: SRPM2} Long-range spectral correlations visible in large $L$ behavior of 
the number variance $\Sigma^2(L)$ for XXZ spin chain \eqref{eq: XXZ} during the MBL transition.
The dashed lines denote predictions of the wSRPMs with parameters from Tab.~\ref{tab}. 
}
\end{figure}

The number variances for the XXZ spin chain \eqref{eq: XXZ} at various disorder strengths $W$
together with the wSRPM {results} from {Sec.~\ref{subsecdiso}}
are presented in Fig.~\ref{fig: SRPM2}. 
Nearly perfect
agreement between the XXZ spin chain data and the predictions of wSRPM 
visible in Fig.~\ref{fig: SRPM1} for $L\in[0,20]$ is lost. Small deviations 
from the linear behavior of the number variance predicted by wSRPM appear at larger scales
{which was {also} indicated by the slight discrepancies between spectral 
compressibility $\chi$ of the data and the prediction of wSRPM}.
There are two distinct regimes. For metallic systems with disorder strengths 
$W \lesssim 2.1$ the number variance obtained from the wSRPM is smaller than 
the result for XXZ spin chain. This indicates that there exists a regime  (for $L\gtrsim 20$)
where the number variance grows faster than linearly which was 
interpreted in \cite{Bertrand16} as a signature of anomalous Thouless energy 
in the system \cite{Braun95}.  We indicate below that this  
behavior of the number variance for large $L$ has to be examined with an uttermost caution. 
The second regime arises as the disorder strength increases above $W \approx 2.5$.
Then, the number variance predicted by wSRPM slightly overestimates the number variance
for the XXZ spin chain. As we have {shown in Fig.~\ref{fig: SRPM3a}.} this effect diminishes as one 
changes the system size from  $K=14$ through $K=16$ to $K=18$ and thus it is likely a finite size
effect. However, we cannot completely exclude the possibility that there are some 
remaining long-range correlations between eigenvalues in the system which are not grasped within the 
wSRPM.

{The simple form of JPDF of wSRPM allows us to get further insight into long-range spectral correlations 
of system in the MBL crossover.}
The situation in which wSRPM accurately reproduces level statistics up to $10$-$20$ level spacings
but underestimates the number variance for $L \gtrsim 20$ is at first sight paradoxical. The wSRPM
incorporates interactions between energy levels only at a finite range {$h=\max\{h_i\}$}. 
In addition, the weaker the correlation between eigenvalues separated by a given distance the bigger {is}
the number
variance at $L$ corresponding to this distance. 
How it is, therefore, possible that wSRPM grasps faithfully the level statistics at the local 
scale but predicts stronger correlations at larger scales as compared to the data for the XXZ spin chain 
while at the same time it does not assume presence of any interactions between energy levels 
beyond the range $h$? It turns out that fluctuations of density of eigenvalues on scales of tens level
spacings increase the number variance at large $L$. This is precisely the moment
in which the unfolding enters the scene as it is the way in which the $\overline \sigma(E)$ 
is defined which determines whether the density fluctuations are incorporated into $\overline \sigma(E)$
resulting in number variance $\Sigma^2(L)$ growing  linearly with $L$ (or, conversely, they are not incorporated and then 
$\Sigma^2(L)$ increases faster than linearly for large $L$).

\begin{figure}
\includegraphics[width=1\columnwidth]{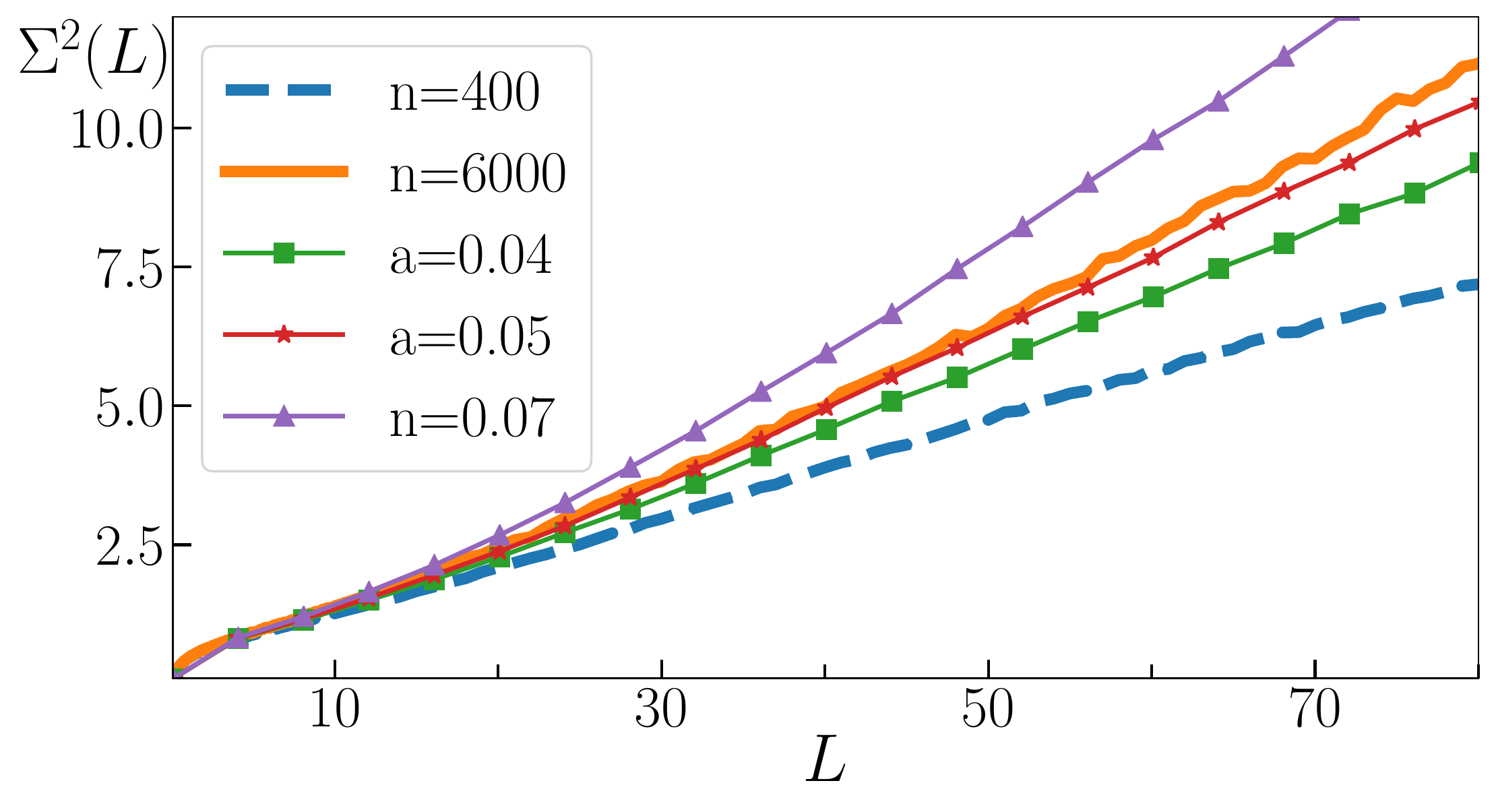}
\caption{ \label{fig: GG1} The number variance $\Sigma^2(L)$ for 
XXZ spin chain \eqref{eq: XXZ} of size $K=18$ for $W=1.5$. The dashed line show result
for local linear unfolding with $n=400$, the solid line for unfolding based on mean density of states \cite{Bertrand16}
with $n=6000$.  The number variance  $\Sigma^2(L)$ obtained after introducing of fluctuations
of density of eigenvalues with parameter $a$ are denoted with lines with markers. }
\end{figure}
The work \cite{Bertrand16} reports  that the number variance grows according to a power-law
$\Sigma^2(L) \propto L^{\gamma}$ with $\gamma > 1$ for large $L$ for the XXZ spin chain \eqref{eq: XXZ} 
deep in the metallic regime where the exponent $\gamma$ acquires values up to $\gamma \approx 1.4$. 
The number variance obtained by us for $W=1.9$ has clearly some region in which it increases faster
than linearly, but such a power-law growth is not observed by us. 
This discrepancy has its root in the unfolding. Unfolding employed in \cite{Bertrand16}
relies on assumption that the shape of mean density of states obtained for the system at given disorder 
strength can be used (after appropriate linear transformations) to unfold large portions of spectrum of 
the system taking $n\approx 6000$ consecutive
energy levels for $K=18$. The fluctuations of density of eigenvalues on the scales of tens or hundreds 
of eigenvalues which are different for different disorder realizations are not incorporated 
in the $\overline \sigma(E)$ as it is determined by the mean density of states in which such fluctuations
are averaged out.

Fig.~\ref{fig: GG1} compares the number variances obtained after the local linear unfolding with $n=400$
consecutive eigenvalues and after the unfolding of \cite{Bertrand16}. The results agree up to $L\approx 15$.
In order to show that the difference between the results stems from the density fluctuations 
we introduce a particular density modulation to the data from the local linear unfolding. Namely, 
the unfolding is modified so that the eigenvalues are mapped via 
\begin{equation}
 E_i \rightarrow \epsilon_i =
\overline \sigma(E_i) + a(E_i - E_C)^2,
\label{eq: unf2}
\end{equation}
where $E_C$ lies in the middle of the energy interval which is unfolded. The $a(E_i - E_C)^2$ term
 mimics the density fluctuations which were not incorporated into $\overline \sigma(E)$,
for $a=0$ \eqref{eq: unf2} reduces to the local linear unfolding \eqref{eq: unf1}. 
 Such a density modulation does not alter $P(s)$ at all, however, it modifies the number variance
 exactly in the manner which allows us to reproduce the result of \cite{Bertrand16} and showing 
 that the density fluctuations are the mechanism which causes the 
 power-law growth of the number variance.

In conclusion, the behavior of the number variance $\Sigma^2(L)$ suggests 
that  long-range spectral correlations might be present in the level statistics of  XXZ spin chain 
during MBL transition. This feature of MBL transition lies beyond 
the scope of wSRPM, however, as we demonstrate{d by examining bosonic and fermionic systems} 
 it is model dependent.
It is not clear whether the unfolding employed in \cite{Bertrand16} is
justified. As we have indicated, it does not take into account variations of density of eigenvalues
at scales of tens and hundreds of level spacings for a given disorder realization. Let us
{emphasize once again}
that the situation {of localization of an interacting system}
differs starkly from the usual RMT where a random matrix depends on number of random 
entries which scales as square of its size whereas the number of random entries 
in the Hamiltonian of the XXZ spin chain scales only as logarithm of the size of the Hilbert space of the system.
Therefore one may expect that while the density fluctuations average out 
for RMT and using Wigner's semi-circle to unfold GOE is a good idea it may not be the case for 
the many body quantum systems which undergo MBL transition.

\section{Conclusions and beyond} 

Analyzing the gap ratios of systems in the crossover between ergodic and many-body localized regimes we have shown 
that a complete information about inter- and intra-sample randomness can be obtained from the  $r_n$ variables.
Distribution $P(r_S)$ of the sample averaged gap ratio $r_S$ provides a suggestive illustration of Griffiths regions 
in the case of random disorder while it shows absence of such rare events in systems with quasiperiodic disorder.
The proposed inter- and intra-sample variances $V_S$ and $V_I$ provide a straightforward method to quantify inter-
and intra-sample randomness. While our analysis provides further insights into role played by Griffiths regions in 
MBL transition, it is conceptually and computationally (involves only eigenvalues) simple and therefore can be
straightforwardly employed in studies of other systems where a transition between integrable and RMT-like regimes occurs.
  
Examining the bulk and the tail of level spacing  distribution together with the number variance 
we have demonstrated that the proposed models of spectral statistics {in} MBL 
{crossover}
\cite{Serbyn16,Shukla16,Bertrand16,Buijsman18} grasp
level statistics {accurately} 
only at the level of few of level spacings.  To reproduce broad distributions 
of the sample averaged gap ratio $r_S$ in the MBL transition we have 
introducted the wSRPM that is a  statistical mixture of
the well known family of short-range plasma models.
The wSRPM describes faithfully the flow of level statistics 
in the whole ergodic to MBL  crossover. 
According to wSRPM {the correlations between} 
eigenvalues  are present only at  a finite
range $h$. In the ergodic phase 
the range $h$ diverges resulting in GOE statistics and as the system flows towards MBL phase
the range of correlations diminishes. At a certain point the interactions become local ($h=1$),  finally in the 
vicinity of MBL phase the level repulsion vanishes ($\beta \rightarrow 0$) resulting in the Poisson statistics.
The wSRPM grasps universal features of level statistics across MBL transition in a variety 
of spin, bosonic and fermionic systems with interactions and random disorder. {The
assumption}  that there are no correlations between eigenvalues 
at ranges larger than $h$ predicts the finite spectral compressibility $\chi$ in the transition.
The latter seems to be approximately true for the studied systems albeit small deviations from the 
linear behavior of the number variance have been noticed. 
This may be either an artifact of the unfolding procedure or could also stem from 
weak long-range 
interactions between energy levels which are model and  system size dependent.

We also considered a weighted ensemble of power law banded random matrices -- see Appendix \ref{sec: wplrbm}.
An appropriate mixture of PLBRM (again necessitated by a broad distribution of gap ratio in physical samples)
seems to be at least competitive with wSRPM leading to small deviations of the fitted
model from the data for XXZ spin chain. Both approaches have their advantages.
While for {SRPMs the eigenvalues} may be generated by brute force Monte Carlo integration 
{of the JPDF}, a softer
semi-analytic approach, working at certain range of eigenvalues interaction, $h$, is possible
following the path shown by Bogomolny and coworkers \cite{Bogomolny01} {as shown in Appendix \ref{AppA}}. 
It provides  expressions for the level spacing distribution $P(s)$ and, more importantly, 
gives analytical formulas for asymptotic behavior of the number variance $\Sigma^2(L)$ as well as for the
tails of $P(s)$. Moreover, the wSRPM gives a concrete microscopic description of correlations between eigenvalues
across the whole MBL crossover and allows us to speculate how the level statistics
evolve in the limit of large system sizes. 

On the other hand that approach
provides us with no clue on the eigenvectors behavior. On the contrary, PLBRM model provides
access to both eigenvalues and eigenvectors by a direct (although costly) diagonalization of
a large number of matrices from the ensemble. The drawback of this approach is that
there are no analytical results for this model at finite $N$ or $\mu \neq 1$ so
a clear picture of correlations between eigenvalues is not available.

 Finally, it is {also} interesting to note that the MBL transition for the quasiperiodic disorder
 case cannot be described by the proposed weighted ensembles. 
 It supports the claim of \cite{Khemani17} that the transitions for
 RD and QPD are of different universality classes.
 The ensemble that reproduces level statistics for QPD in MBL transition is yet to be identified. 
 
\section*{Acknowledgments}
We acknowledge fruitful and enlightening discussions with D. Delande as well as exchanges on unfolding 
procedures with A. M. Garcia-Garcia and M. Sieber.
This work was performed with the support of EU via Horizon2020 FET project QUIC (nr.~641122). 
Numerical results were obtained with the help of PL-Grid Infrastructure. We acknowledge support of the 
National Science Centre (PL) via project No.2015/19/B/ST2/01028 (P.S.), {No.2018/28/T/ST2/00401 
(Etiuda scholarship -- P.S.)} and the  QuantERA
programme No. 2017/25/Z/ST2/03029 (J.Z.). 

\newcommand{\snum}{A}
\setcounter{section}{0}
\renewcommand{\theequation}{\snum.\arabic{equation}}

\setcounter{equation}{0}

\section{Appendix: analytical expressions for short-range plasma model}
\label{AppA}
{
The following quantities are needed in order to construct wSRPM: level spacing distributions $P^{\beta}_h(s)$,
number variances $\Sigma^2_{\beta,h}(L)$, the sample gap ratio distribution $P^{\beta}_h(r_S)$ of the individual SRPMs used 
in the wSRPM.
Semi-analytic expression for most of those quantities have been given in Ref. \cite{Bogomolny01}. We comment here on how 
they can be used in the context of wSRPM.
}

{
Let us start with $h=1$. The distribution of level spacing between  $n$-th neighboring eigenvalues
reads
\begin{equation}
 P^{\beta}_{h=1}(n,s) = \frac{(\beta+1)^{(n+1)(\beta+1)} }{ \Gamma((\beta+1)(n+1))}s^{\beta+n(\beta+1)}\mathrm{e}^{-(\beta+1)s},
 \label{SRPM_P1}
\end{equation}
where $\beta \in [0,1]$, for $n=0$ this distribution reduces to the usual level spacing distributions. For $n \geq 0$,
the formula \eqref{SRPM_P1} can be used to obtain the number variance according to the following general
expression
\begin{equation}
 \Sigma^2(L) = L - 2 \int_0^L \mathrm{d}s (L-s) \left( 1- \sum_{n=0}^{\infty} P(n,s) \right).
 \label{SRPM_P2}
\end{equation}
From our Monte Carlo evaluation of JPDF \eqref{eq: SRPM} we know that the $P^{\beta}_{h}(r_S)$ distributions
(for range $h$ and $\beta$ relevant in applications of our wSRPM)
are Gaussian functions, localized around certain values $\overline r^{\beta}_{h}$ with standard deviation $\sigma\approx0.01557$
which changes only very slightly for various $h$ and $\beta$.
Therefore, instead of deriving the full $P^{\beta}_{h=1}(r_S)$ distribution, we simply use 
the distribution of gap ratio to calculate $\overline r^{\beta}_{h=1}$ and use the Gaussian approximation for 
the full distribution $P^{\beta}_{h=1}(r_S)$.
The gap ratio distribution $P^{\beta}_{h=1}(r)$
reads \cite{Atas13b}
\begin{equation}
 P^{\beta}_{h=1}(r)=\frac{\Gamma(2\beta+2) \Gamma^2(\beta+2)}{(\beta+1)^2 \Gamma^4(\beta+1)} \frac{2 r^{\beta}}{(1+r)^{2\beta+2}},
\end{equation}
from which we get $\overline r^{\beta}_{h=1}$ as the first moment of $P^{\beta}_{h=1}(r)$ distribution.
}

{ For $h>1$ the only interesting case for us is $\beta=1$. Then, the level spacing distribution is given by \eqref{eq: PS1}. 
In order to determine the polynomial $W(s)$ as well as the distributions $ P^{\beta}_{h}(n,s)$ for  $h>1$ and $n>0$  one needs to
solve the integral equation\cite{Bogomolny01}
\begin{eqnarray}
\nonumber
 \int_0^{\infty}\mathrm{d}\xi_h \mathrm{e}^{-\xi_h} \xi_h
 (\xi_h+\xi_{h-1})\ldots(\xi_h+\ldots+\xi_1) &
 \psi_j({\xi_2,\ldots,\xi_h}) \\
 = \lambda_j \psi_j(\xi_1,\ldots,\xi_{h-1}),
 \label{inteq1}
\end{eqnarray}
for function  $\psi_j(\xi_1,\ldots,\xi_{k-1})$. The equation can be solved by a polynomial ansatz
\begin{eqnarray}
\nonumber
\psi_j(\xi_1,\ldots,\xi_{h-1}) = \\ 
\sum_{i_1=0}^{1}\sum_{i_2=0}^{3} \ldots   \sum_{i_{h-1}=0}^{(h-1)h/2} &
a_{i_1i_2\ldots i_{h-1}} & \xi^{i_1}_1 \xi^{i_2}_2 \ldots\xi^{i_{h-1}}_{h-1},
\end{eqnarray}
which reduces \eqref{inteq1} to an eigenproblem for a matrix of dimension $D_h=\prod_{k=1}^{h-1}i_k$. 
Solving the eigenproblem, the eigenfunctions $\psi_j(\xi_1,\ldots,\xi_{h-1})$ can be used to find 
$ P^{\beta}_{h}(n,s)$ for $h>1$ and $n\geq0$ as well as $ P^{\beta}_{h}(r)$ which is then used to determine
$P^{\beta}_{h}(r_S)$ as a Gaussian function with standard deviation $\sigma = 0.01557$ centered around 
$\overline r^{\beta}_{h}$. The level spacing distribution is determined as $ P^{\beta}_{h}(s) =  P^{\beta}_{h}(n=0,s)$,
whereas $ P^{\beta}_{h}(n=0,s)$ for $5>n\geq0$ are used to find the number variance  $\Sigma^2_{\beta,h}(L)$ for $L<3$.
For larger values $L \geq 3$ the asymptotic form \eqref{eq: PS2} of the number variance is used.
}

{Unfortunately, the dimension $D_h$ grows exponentially with $h$ which makes the 
semi-analytic approach feasible only for $h<6$  which incidentally   covers all the SRPMs used in \eqref{c12}. The 
coefficients which determine $P^{\beta}_{h}(n,s)$ are gathered in Tab.~\ref{tabA1}, \ref{tabA2},
\ref{tabA3},\ref{tabA4}.
The $P^{\beta=1}_{h}(r_S)$ distributions are approximated as Gaussian distributions with 
standard deviation $\sigma = 0.01557$ centered around 
the semi-analytically obtained values of $\overline r^{\beta=1}_{h}$ which are $0.5155,0.5206,0.5231,0.5246$
for $h=2,3,4,5$ respectively.
}

%


\begin{table*}
\begin{tabular}{S | S S S S S } \toprule
      & {$\quad  n=0$} & {$\quad \quad  n=1$} & {$\quad \quad \quad  n=2$} & {$\quad \quad \quad  n=3$} & {$\quad \quad \quad  n=4$}\\ \midrule
    {$w_1$}  & 2.4773 &  2.50542 & 0.325782  & 0.0160653 &  1.9203 $\cdot 10^{-4}$ \\
    {$w_2$}  & 6.06811 & 3.0685 &   0.243833 & 6.51294$\cdot 10^{-3}$ &  8.2295 $\cdot 10^{-5}$  \\
    {$w_3$}  & 3.71594 & 0.751625 &   4.0723$\cdot 10^{-2}$ &  6.0483 $\cdot 10^{-4}$ & 8.2297 $\cdot 10^{-6}$ \\ \bottomrule
 \end{tabular}
 \caption{ \label{tabA1} Coefficients, $h=2$, 
 $P(n,s)=\exp(-3s)s^{3n} \sum_{j}w_j s^{j}$.
 } 
\end{table*}







\begin{table*}
\begin{tabular}{S | S S S S S } \toprule
      & {$\quad  n=0$} & {$\quad \quad  n=1$} & {$\quad \quad \quad  n=2$} & {$\quad \quad \quad  n=3$} & {$\quad \quad \quad  n=4$}\\ \midrule
    {$w_1$}  & 2.13422  & 1.13398                  &  0.14101               &  3.03041$\cdot 10^{-3}$  & 1.7921$\cdot 10^{-5}$\\
    {$w_2$}  & 7.81963  & 3.63343                  &  0.29749               &  4.48832$\cdot 10^{-3}$  &  2.0107$\cdot 10^{-5}$  \\
    {$w_3$}  & 11.6945  & 5.01886                  &  0.26416               &  2.81378$\cdot 10^{-3}$ & 9.7956$\cdot 10^{-6}$ \\
    {$w_4$}  & 9.07429  & 3.71122                  &  0.12338               &  9.58731$\cdot 10^{-4}$ & 2.6381$\cdot 10^{-6}$ \\ 
    {$w_5$}  & 3.68554  & 1.50368                  &  3.2011$\cdot 10^{-2}$ &  1.87245$\cdot 10^{-4}$ & 4.1257$\cdot 10^{-7}$ \\ 
    {$w_6$}  & 0.62587  & 0.310017                 &  4.455 $\cdot 10^{-3}$ &  2.00989$\cdot 10^{-5}$ & 3.5835$\cdot 10^{-8}$  \\
    {$w_7$}  & 0        & 2.59412 $\cdot 10^{-2}$  &  2.6414$\cdot 10^{-4}$ &  9.37094$\cdot 10^{-7}$ & 1.3637$\cdot 10^{-9}$  \\
    \bottomrule
 \end{tabular}
 \caption{ \label{tabA2} {Coefficients, $h=3$,  $P(n,s)=\exp(-4s)s^{4n-1+\delta_{0,n}} \sum_{j}w_j s^{j}$, 
 where $\delta_{i,j}$ denotes the Kronecker delta.}
 } 
\end{table*}

\begin{table*}
\begin{tabular}{S | S S S S } \toprule
      & {$\quad  n=0$} & {$\quad \quad  n=1$} & {$\quad \quad \quad  n=2$} & {$\quad \quad \quad  n=3$} \\ \midrule
    {$w_1$}    & 1.9869                  & 0.810475                &  0.0483296              &  7.29136$\cdot 10^{-4}$  \\
    {$w_2$}    & 9.44665                 & 3.61582                 &  0.190586               &  2.12001$\cdot 10^{-3}$    \\
    {$w_3$}    & 20.1982                 & 7.54195                 &   0.351477              &  2.85649$\cdot 10^{-3}$  \\
    {$w_4$}    & 25.6905                 & 9.67193                 &  0.399671               &  2.35667$\cdot 10^{-3}$  \\ 
    {$w_5$}    & 21.5383                 & 8.44298                 &   0.310511              &  1.32441$\cdot 10^{-3}$ \\ 
    {$w_6$}    & 12.39                   & 5.24733                 &  0.172672               &  5.34180$\cdot 10^{-4}$   \\
    {$w_7$}    & 4.90293                 & 2.35889                 &  7.01559$\cdot 10^{-2}$ &  1.58784$\cdot 10^{-4}$   \\
    {$w_8$}    & 1.28897                 & 0.763259                &   2.0934$\cdot 10^{-2}$ &  3.51389$\cdot 10^{-5}$   \\
    {$w_9$}    & 0.204419                & 0.17333                 &  4.55485$\cdot 10^{-3}$ &  5.76574$\cdot 10^{-6}$   \\
    {$w_{10}$} & 0.0148948               & 2.61995$\cdot 10^{-2}$  &  7.06631$\cdot 10^{-4}$ &  6.86517$\cdot 10^{-7}$   \\
    {$w_{11}$} & 0                       & 2.36573$\cdot 10^{-3}$  &  7.44424$\cdot 10^{-5}$ &  5.64602$\cdot 10^{-8}$   \\
    {$w_{12}$} & 0                       & 9.63045$\cdot 10^{-5}$  &  4.79282$\cdot 10^{-6}$ &  2.87566$\cdot 10^{-9}$   \\
    {$w_{13}$} & 0                       & 0                       &  1.42564$\cdot 10^{-7}$ &  6.85001$\cdot 10^{-11}$   \\
    \bottomrule
 \end{tabular}
 \caption{ \label{tabA3} {Coefficients, $h=4$,  $P(n,s)=\exp(-5s)s^{(n+1)(n+4)/2-2} \sum_{j}w_j s^{j}$.}
 } 
\end{table*}

\begin{table*}
\begin{tabular}{S | S S S S } \toprule
      & {$\quad  n=0$} & {$\quad \quad  n=1$} & {$\quad \quad \quad  n=2$} & {$\quad \quad \quad  n=3$} \\ \midrule
    {$w_1$}    & 1.90607                 & 0.6725112              & 2.816938$\cdot 10^{-2}$&  1.90849$\cdot 10^{-4}$  \\
    {$w_2$}    & 11.0655                 & 3.757313               & 0.1474950               & 8.91886$\cdot 10^{-4}$    \\
    {$w_3$}    & 30.0426                 & 10.08246               & 0.37248642              & 2.00465$\cdot 10^{-3}$  \\
    {$w_4$}    & 50.783                  & 17.24174               & 0.60284124              & 2.87998$\cdot 10^{-3}$  \\ 
    {$w_5$}    & 59.9035                 & 21.02683               & 0.70045560              & 2.96413$\cdot 10^{-3}$ \\ 
    {$w_6$}    & 52.2919                 & 19.39166               & 0.62020              & 2.32137$\cdot 10^{-3}$   \\
    {$w_7$}    & 34.9026                 & 13.99334               & 0.43344442               & 1.43410$\cdot 10^{-3}$   \\
    {$w_8$}    & 18.1222                 & 8.06393                & 0.2443500             & 7.15049$\cdot 10^{-4}$   \\
    {$w_9$}    & 7.36415                 & 3.75359                & 0.11260234 & 2.92059$\cdot 10^{-4}$   \\
    {$w_{10}$} & 2.33283                 & 1.41789                & 4.27332$\cdot 10^{-2}$  &  9.8650$\cdot 10^{-5}$   \\
    {$w_{11}$} & 0.567501                & 0.4341132              & 1.339465$\cdot 10^{-2}$ &  2.77077$\cdot 10^{-5}$   \\
    {$w_{12}$} & 0.102858                & 0.1069601              & 3.464275$\cdot 10^{-3}$ &  6.48527$\cdot 10^{-6}$   \\
    {$w_{13}$} & 4.05358$\cdot 10^{-5}$  & 2.09130$\cdot 10^{-2}$ & 7.357053$\cdot 10^{-4 }$&  1.26372$\cdot 10^{-6}$   \\
    {$w_{14}$} & 0                       & 3.17047$\cdot 10^{-4}$ & 1.271140$\cdot 10^{-4 }$&  2.04088$\cdot 10^{-7}$   \\
    {$w_{15}$} & 0                       & 3.59091$\cdot 10^{-4}$ & 1.760258$\cdot 10^{-5 }$&  2.70805$\cdot 10^{-8}$   \\
    {$w_{16}$} & 0                       & 2.85564$\cdot 10^{-6}$ & 1.908644$\cdot 10^{-6 }$&  2.908829$\cdot 10^{-9}$  \\
    {$w_{17}$} & 0                       & 1.41936$\cdot 10^{-6}$ & 1.561970$\cdot 10^{-7 }$ &  2.44768$\cdot 10^{-10}$  \\
    {$w_{18}$} & 0                       & 3.30973$\cdot 10^{-8}$ & 9.07438$\cdot 10^{-9 }$ & 1.62366$\cdot 10^{-11}$   \\
    {$w_{19}$} & 0                       & 0                      & 3.33532$\cdot 10^{-10}$&   5.17973$\cdot 10^{-13}$   \\
    {$w_{20}$} & 0                       & 0                      & 5.82950$\cdot 10^{-12}$  &  1.11711$\cdot 10^{-12}$   \\
    \bottomrule
 \end{tabular}
 \caption{ \label{tabA4} {Coefficients, $h=5$,  $P(n,s)=\exp(-6s)s^{(n+1)(n+4)/2-2} \sum_{j}w_j s^{j}$.}
 } 
\end{table*}

\section{Appendix: Weighted power-law random banded matrices}
\label{sec: wplrbm}
The wSRPM describes faithfully level statistics in MBL transition. However, it 
provides no information on
 properties of eigenstates.
One particularly interesting property is multifractality of matrix elements of local 
operators \cite{Monthus16,Serbyn17} in such states. Therefore, an identification of a random matrix model
which could provide some information about eigenvectors in MBL transition can be productive. 

In this Appendix we examine an ensemble of power-law random banded matrices (PLRBM)
\cite{Mirlin96,Evers08} which is the ensemble of
$N\times N$ symmetric real matrices with matrix  elements $H_{ij}$ being
independent random Gaussian variables with                                                                             
\begin{equation}                                                                                              
\langle H_{ij} \rangle = 0 \,\,\,\,\, \mathrm {and} \,\,\,\,\,\,\,                  
\langle H_{ij}^2 \rangle = (1+\delta_{ij})\left( 1+ \left(|i-j|/B\right)^{2\mu} \right)^{-1}.                
\label{eq: CBRM}                                                                                             
\end{equation}  
This ensemble interpolates between GOE statistics for $B  \gg 1 $, $\mu < 1$
and PS statistics which arises for $\mu>1$ in $N\rightarrow \infty$ limit. 
In the special case of $\mu=1$ and large $B$ the model can be solved by a mapping onto an effective
$\sigma$ model \cite{Mirlin00}. Numerical calculations of level statistics of the PLRBM
model at the critical line 
$\mu=1$ were carried out  in \cite{Varga00,Ndawana03}.
\begin{figure}
\includegraphics[width=1\columnwidth]{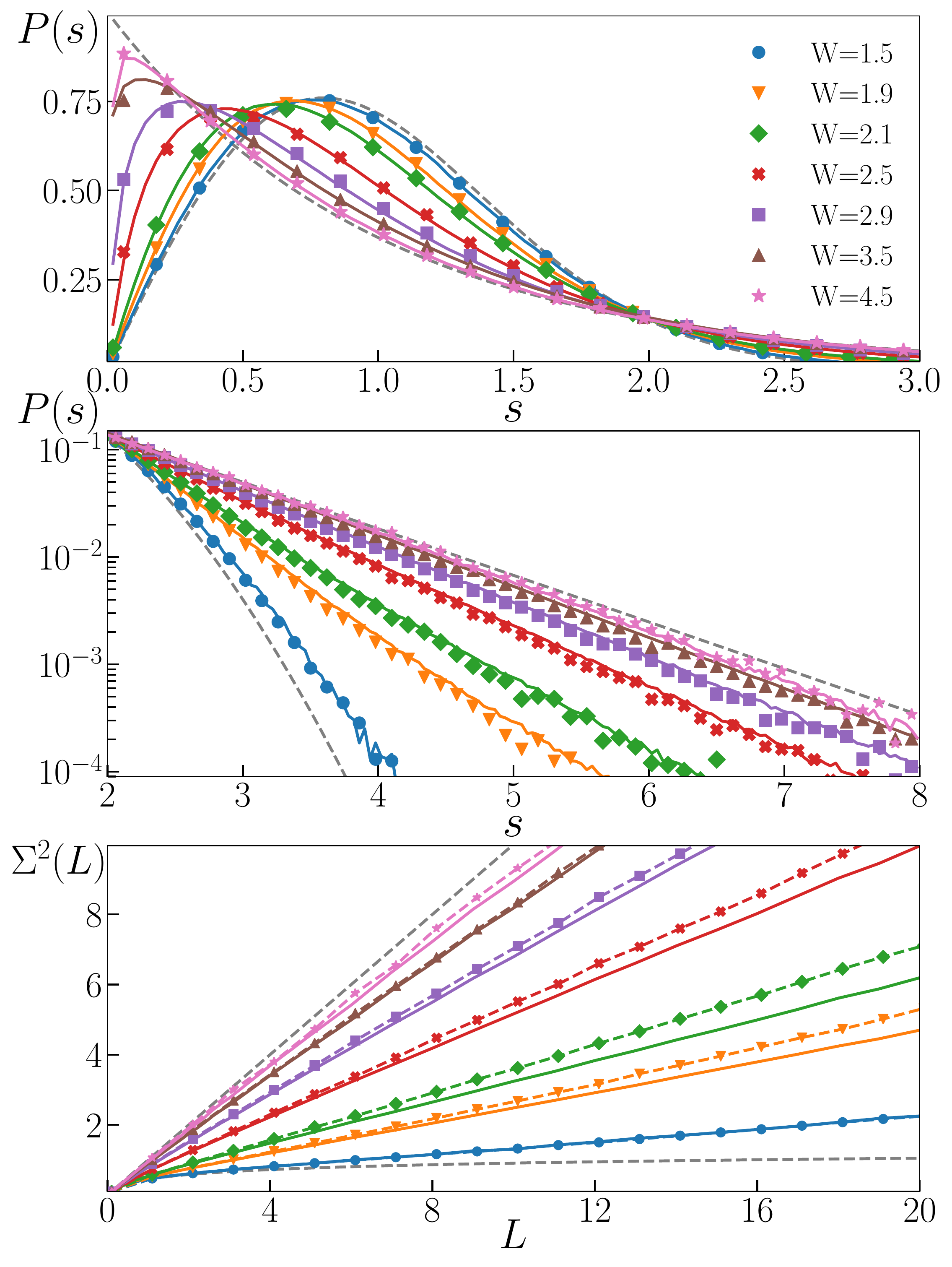}
\caption{ \label{fig: PLBM3} Distributions $P(r_S)$ across the MBL transition 
(denoted by markers) with fits from the weighted PLRBM model.}
\end{figure} 
\begin{figure}
\includegraphics[width=1\columnwidth]{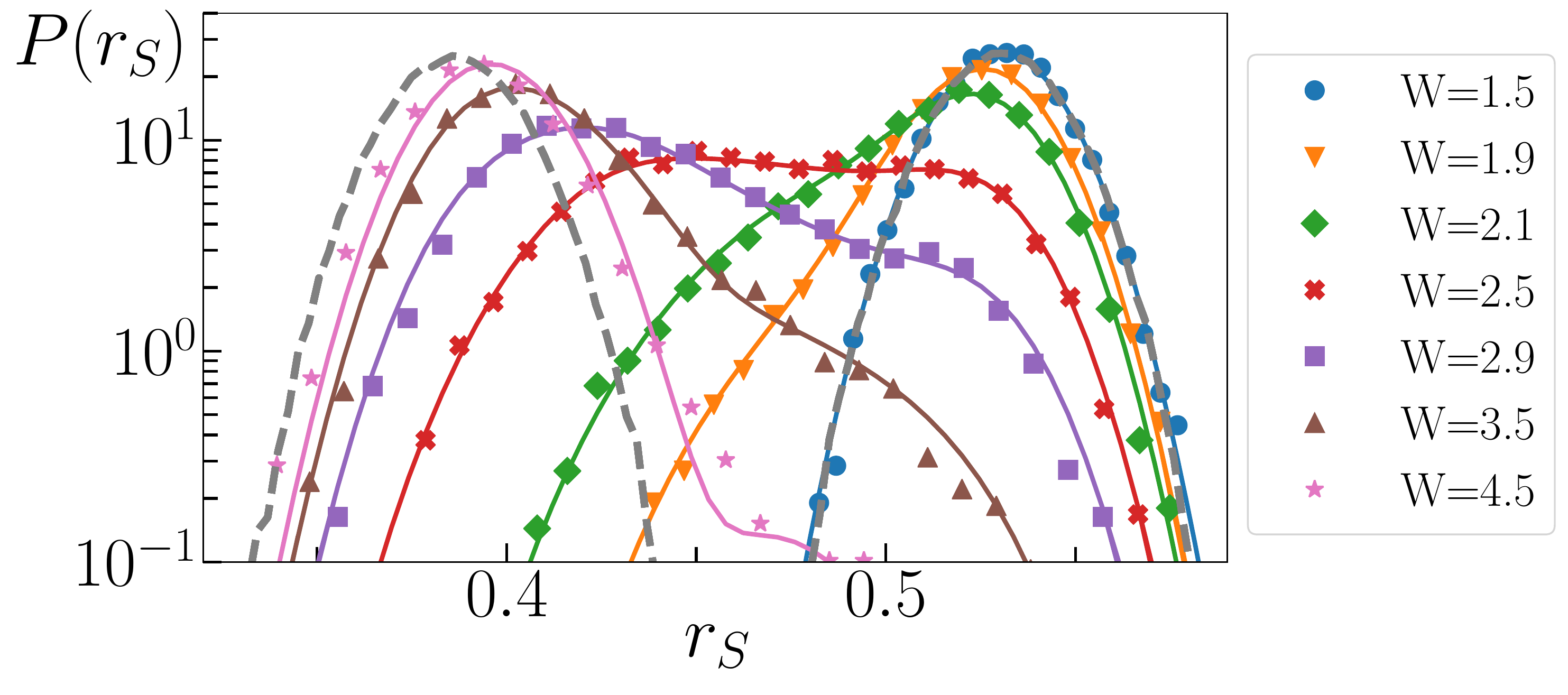}
\caption{ \label{fig: PLBM2} Level statistics of wPLRBM model fitted
to data for XXZ spin chain \eqref{eq: XXZ} with
appropriate $r_S$ filtering -- details in Tab.~\ref{tab2}.
}
\end{figure} 
\begin{figure}
\includegraphics[width=1\columnwidth]{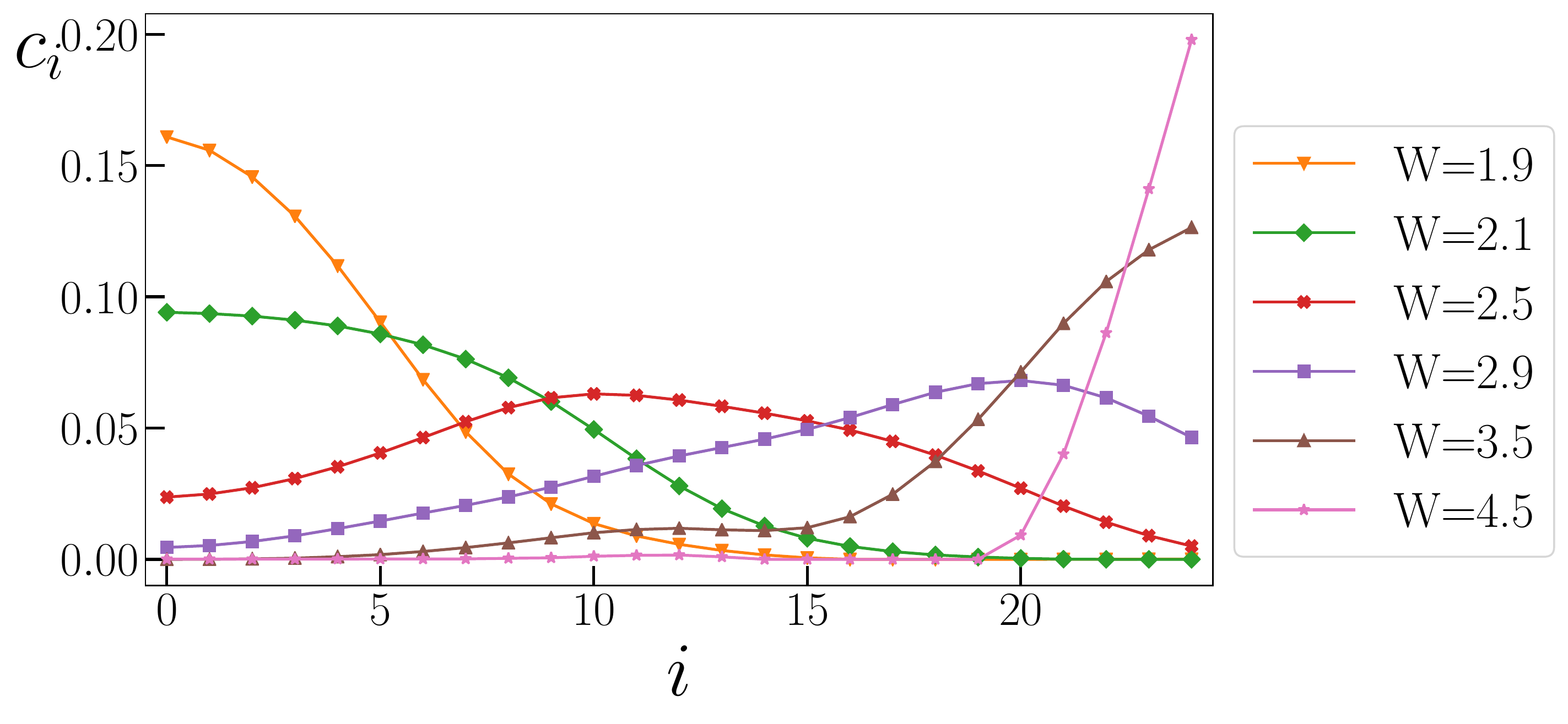}
\caption{ \label{fig: ci3a} {Coefficients $c_i$ for the wPLRBM model
for XXZ spin chain across the ergodic MBL crossover. 
} }
\end{figure}

{We consider PLRBM of size $N=1000$, accumulating $10000$ matrices for each set of parameters $(\mu, B)$.
Let us note that the exact values of the $(\mu, B)$ coefficients are strongly dependent on size $N$ of matrix from 
PLRBM. With growing $N$ a flow of level statistics in this model occurs -- points $(\mu,B)$ with $\mu<1$
correspond to statistics closer and closer to GOE and analogously -- for $\mu>1$ statistics flow towards PS.
Calculating the $P(r_S)$ distribution for PLRBM model we have verified that 
$P(r_S)$ remains Gaussian in large region of parameter
space $(\mu,B)$. Moreover, there exist a region of parameters for which 
the level spacing distributions $P^{\mu}_{B}(s)$ decay exponentially and the number variance 
$\Sigma^2_{\mu,B}(L)$ is asymptotically linear. Therefore, a similar extension as in the case of SRPM 
can be proposed in which the inter-sample randomness encoded in the $P(r_S)$ distribution 
is mimicked by considering a mixture of PLRBM with various $\mu_i$ and $B_i$ to describe the level 
statistics in given point of MBL transition. 
{More precisely, the selected set of PLRBMs consists of models with $B_i=0.35$ and varying 
$\mu_i\in\{0.75+i*0.025\}_{i=0,1,...23,22,24,26,28,32,39,49}$. The corresponding 
weight coefficients are obtained by minimizing $\chi^2$ as in \eqref{c13} with results shown in
Fig.~\ref{fig: ci3a}. {The $W=1.5$ was fitted with single PLRBM model with $B=0.35$, $\mu=0.7$.} }

 Let us note that this model gives very good agreement 
at the level of ten{s} of level spacings.  {Tails of} the level spacing
distributions as well as the number variance obtained from weighted PLRBM model are compatible with
the XXZ spin chain data.  
Certain deviations  are
visible in the spectral compressibility $\chi$ {at larger $L$}.

The PLRBM model was introduced as a model for studies of critical properties of Anderson localization. 
In its direct interpretation the model \eqref{eq: CBRM} describes a single particle on 
one dimensional sample with disorder and with long-range hopping -- tunneling amplitude decays
according to a power-law with distance.
Our results show that the PLRBM can be used also in MBL transition 
provided {the} weighted mixture of matrices is considered.

Such an ensemble needs to be introduced to mimic the large inter-sample randomness in the MBL crossover,
which is a specific feature of localization of an interacting system, whose exponentially large in $L$
Hamiltonian matrix depends only on $L$ random variables.

One way of 
interpreting this result is that MBL can be thought of as a single particle 
localization in a 'Fock-space lattice' with complex geometry \cite{Welsh18,Welsh18b} (reflecting
the quantum many-body character of the phenomenon).
Another approach is to view wPLRBM as the Hamiltonian of the system at late stages
of diagonalization flow 
\cite{Rademaker16,Monthus16b,Thomson18} so that the diagonal entries represent random eigenergies associated
with soon-to-be LIOMs and the quickly decaying off-diagonal elements account for still present interactions which
become weaker and weaker close to the  MBL phase.

If the latter is true, then to get the multifractal properties of matrix elements of local operators
\cite{Monthus16,Serbyn17} one has to know transformation between the $\sigma_i^z$ eigenbasis (in which 
the Hamiltonian matrix is straightforwardly computed) and 
the basis in which the Hamiltonian becomes the banded matrix. This would also be the basis in which 
an interesting relation between the multifractal dimension $D_1$ and 
the spectral compressibility $\chi$ holds. This is beyond the scope of the present paper.

%

  \end{document}